\documentclass[a4paper,11pt]{article}
\usepackage{jheppub}
\usepackage[T1]{fontenc}

\def\Nc{\mathrm{N_c}}
\def\CF{\mathrm{C_F}}
\def\CA{\mathrm{C_A}}
\renewcommand{\d}{\mathrm{d}}

\title{\textbf{Resummed jet mass distribution with trimming in Z+jet events at the LHC}}

\author[a]{\textbf{Safa Gaid},}
\author[a,1]{\textbf{Yazid Delenda}\note{Corresponding author.}}
\author[b,c]{\textbf{and Rachik Soualah}}

\affiliation[a]{Laboratoire de Physique des Rayonnements et de leurs Interactions avec la Mati\`{e}re,\\
D\'{e}partement de Physique, Facult\'{e} des Sciences de la Mati\`{e}re, Universit\'{e} de Batna-1,\\
 Batna 05000, Algeria}
\affiliation[b]{Department of Physics, Khalifa University of Science and Technology,\\
P.O. Box 127788, Abu Dhabi, United Arab Emirates}
\affiliation[c]{The International Center for Theoretical Physics (ICTP),\\
Strada Costiera 11, Trieste I-34151, Italy}

\emailAdd{safa.gaid@univ-batna.dz}
\emailAdd{yazid.delenda@univ-batna.dz}
\emailAdd{rachik.soualah@ku.ac.ae}

\abstract{In this paper, we calculate the resummed jet mass distribution up to next-to-leading logarithmic accuracy for jets defined with the trimming groomer in Z+jet events at the LHC. We compute at fixed order and to all orders the large logarithms, both in the jet mass and $z_{\mathrm{cut}}$ trimming variable, including non-global and clustering logarithms. We conduct phenomenological studies at $\mathcal{O}(\alpha_s)$, validating our results by comparing them with the fixed-order Monte Carlo program \texttt{MCFM}, which provides NLO predictions for hadron colliders. The resummation of non-global logarithms is estimated in the large-$\Nc$ limit. Our resummed result is compared to Monte Carlo parton shower simulations from \texttt{Pythia 8}, \texttt{Herwig ++} and \texttt{Sherpa 2} event generators.}

\keywords{Jets and Jet Substructure, Resummation}

\begin{document}
\maketitle
\flushbottom

\section{Introduction}

Jets are copiously produced at the CERN-Large Hadron Collider (LHC), predominantly arising from the hard QCD dynamics of partons and their associated soft emissions, as well as contributions from the underlying event and pileup interactions which further complicate the jet environment. Jets also emerge from the hadronic decay products of heavy particles, particularly top quarks, weak bosons, and the Higgs boson - particles that are key targets of study at collider experiments. Moreover, potential new physics signals, such as Dark Matter candidates from different physics scenarios, might also decays into jets. The decay products of a heavy particle with mass $M$ and transverse momentum $p_t$ typically cluster within a radius of $R\sim 2M/p_t$. In the extreme regions of phase space reached by the high-energy proton-proton collisions at the LHC, these heavy particles can acquire significant Lorentz boosts, leading to collimated decay products that eventually manifest as a single large-radius fat jet.

Distinguishing such electroweak signal events from the overwhelming QCD background is a major challenge. Jet substructure techniques have emerged as powerful tools to address this, enabling more precise discrimination between different types of jets \cite{Abdesselam:2010pt,Altheimer:2012mn,Altheimer:2013yza,Adams:2015hiv}. These jet grooming techniques selectively remove soft wide-angle radiation within jets that is less likely to have originated from the hard parton that initiated the jet, thereby enabling more accurate and cleaner studies of jet properties. Consequently, jet grooming has proven invaluable in mitigating contamination from pileup interactions and the underlying event, a capability that will be crucial as the LHC moves towards its High-Luminosity phase (HL-LHC).

The application of jet substructure techniques is widespread, extending to various new physics searches conducted by the CMS and ATLAS experiments at the LHC — see for example refs. \cite{CMS:2014hvu,CMS:2022pjv,ATLAS:2015xom}. Beyond searches for new physics, these techniques have demonstrated promise for precision QCD measurements, such as determining the strong coupling constant $\alpha_s$ \cite{Marzani:2019evv,Hannesdottir:2022rsl}. Popular grooming algorithms include trimming \cite{Krohn:2009th}, pruning \cite{Ellis:2009su,Ellis:2009me}, filtering/mass drop tagger (MDT) \cite{Butterworth:2008iy}, modified MDT (mMDT) \cite{Dasgupta:2013ihk}, and soft drop \cite{Larkoski:2014wba}. Comprehensive reviews on jet substructure can be found in refs. \cite{Larkoski:2017jix,Kogler:2018hem,Marzani:2019hun}.

Among the various observables used in these studies, the invariant mass of the jet stands out as particularly important. The jet mass plays a crucial role in the said disentanglement of electroweak signals from QCD backgrounds and provides insight into perturbative and non-perturbative QCD dynamics. Both the ATLAS and CMS collaborations have conducted experimental studies on the groomed jet mass distribution in a variety of final states, such as dijet events, single jets associated with a vector boson, and boosted $W/Z$ and top production \cite{CMS:2013kfv,CMS:2017pcy,CMS:2018vzn,CMS:2021iwu,ATLAS:2013bqs,ATLAS:2017zda,ATLAS:2019mgf}.

The jet mass observable has attracted substantial interest due to its sensitivity to various QCD dynamics. The distribution of this observable is characterized by large logarithms of the ratio of the jet mass $m_j$ to its transverse momentum $p_t$. At next-to-leading logarithmic (NLL) accuracy, all the leading logarithms $\alpha_s^n \ln ^{n+1} (m_j/p_t)$ and next-to-leading logarithms $\alpha_s^n \ln ^{n} (m_j/p_t)$ in the exponent of the cumulative distribution are resummed to all orders. Studies on the jet mass distribution for ungroomed jets using anti-$k_t$ \cite{Cacciari:2008gp} and $k_t$ \cite{Catani:1993hr,Ellis:1993tq} clustering in vector-boson+jet final states have been conducted in refs. \cite{Dasgupta:2012hg,Ziani:2021dxr}. Additionally, the first analytical calculations of the jet mass distribution using various substructure methods were performed in refs. \cite{Dasgupta:2013ihk,Dasgupta:2013via} for the $e^+e^-$ annihilation to dijets.

A significant focus of theoretical study has been on Non-Global Logarithms (NGLs) \cite{Dasgupta:2001sh,Dasgupta:2002bw} and Clustering Logarithms (CLs) \cite{Banfi:2005gj,Delenda:2006nf}, which appear in the jet mass spectrum for certain grooming algorithms. NGLs arise from the correlated emission of soft gluons, while CLs originate from primary emissions directly from the hard partons. Resumming these logarithms to all orders is complex and often requires numerical Monte Carlo approaches in the large-$\Nc$ approximation, where $\Nc$ is the number of QCD quark colors. These logarithms are present in the jet mass distribution for groomers like trimming and pruning but are absent in others, such as mMDT \cite{Dasgupta:2013ihk,Dasgupta:2013via}. As a result, this has led to the popularity of the mMDT algorithm in jet mass studies \cite{Marzani:2017mva,Marzani:2017kqd,Kang:2018vgn,Anderle:2020mxj,Caletti:2021oor,Reichelt:2021svh,Dasgupta:2022fim}, as it simplifies resummation to NNLL accuracy (and perhaps beyond) using either the soft collinear effective theory (SCET) approach or ordinary QCD perturbation theory.

In this paper, we focus on the NLL resummation of the jet mass distribution in Z+jet final states at the LHC proton-proton collisions for jets defined using the trimming algorithm. This groomer, frequently used by the ATLAS experiment, presents unique challenges, particularly in dealing with the emergence of NGLs and CLs. We perform a fixed-order calculation of the trimmed distribution at leading order and compare our results with those obtained from the Monte Carlo program \texttt{MCFM 9.1} \cite{Campbell:2019dru}, finding agreement. By subtracting the large logarithms from the \texttt{MCFM} distribution, the resulting constant at small jet mass values is used to extend the accuracy of the perturbative jet mass distribution to NLL', accounting for all logarithms of the form $\alpha_s^n\ln^m (m_j/p_t)$, with $m=\{2n, 2n-1, 2n-2\}$ 
in the order-by-order expansion of the cumulative distribution \cite{Banfi:2010xy}.

At next-to-leading order ($\mathcal{O}(\alpha_s^2)$), we calculate the contributions of NGLs and CLs to the trimmed distribution and assess their impact. While CLs are negligible, NGLs remain moderate. We then use a Monte Carlo program \cite{Dasgupta:2001sh,Dasgupta:2002bw} to resum the NGLs at leading-color approximation. Next, we perform a numerical convolution of the resummed global and non-global form factors with the Born cross-section using the \texttt{MadGraph5\_aMC@NLO} \cite{Maltoni:2002qb,Alwall:2014hca} event generator. Our numerical results for the NLL distribution, with and without trimming, are presented, highlighting the impact of trimming on the distribution. Finally, we compare the resummed results with parton shower simulations from \texttt{Pythia 8.3} \cite{Bierlich:2022pfr,Alwall:2008qv}, \texttt{Herwig ++} \cite{Bahr:2008pv,Bellm:2015jjp}, and \texttt{Sherpa 2.2} \cite{Sherpa:2019gpd}.

This paper is organized as follows. In Section 2, we define the kinematics of the process $pp \to Z$+jet and introduce the jet mass observable. We also provide a brief review of the clustering algorithms used and present the general expression for the jet mass distribution as a convolution of the Born cross-section with the resummed global and non-global form factors. In Section 3, we perform a fixed-order calculation of this distribution and compare the results with those obtained using \texttt{MCFM}. In Section 4, we calculate NGLs and CLs at $\mathcal{O}(\alpha_s^2)$ and resum the NGLs to all orders. Section 5 presents our numerical results, comparing them to various simulations of Monte Carlo event generators. Finally, in Section 6, we summarize our findings and conclusions. The notations from refs. \cite{Banfi:2004yd,Dasgupta:2012hg,Ziani:2021dxr} are used throughout this paper.

\section{Jet mass distribution}

\subsection{Process kinematics}

We examine the process of production of a Z boson and a hard jet in $pp$ collisions at a center-of-mass energy $\sqrt{s}$. The $Z$ plus jet final state has garnered significant interest due to its crucial role in various phenomenological studies. These include its presence as an important background in many processes, its clean experimental signature from Z decays, and its use in probing parton distribution functions and the strong coupling constant. One notable example of such studies is the application of the Transverse Momentum Dependent (TMD) resummation formalism to some observable distributions in this process \cite{Sun:2018icb}.

To achieve Single Logarithmic (SL) accuracy in resummation, we assume the emitted gluons are strongly ordered in transverse momentum $k_{ti}$, such that at the $n^\mathrm{th}$ order: $k_{tn} \ll \cdots \ll k_{t2} \ll k_{t1} \ll p_t$, where $p_t$ is the jet's transverse momentum. In this regime, the momenta of the incoming partons $(a)$ and $(b)$, carrying momentum fractions $x_a$ and $x_b$ of the colliding protons, along with the outgoing hard parton $(j)$ and soft emissions $(i)$, are expressed as:
\begin{subequations}
\begin{align}\label{eq:mom}
p_a&=x_a\,\frac{\sqrt{s}}{2}\left(1,0,0,1\right),\\
p_b&=x_b\,\frac{\sqrt{s}}{2}\left(1,0,0,-1\right),\\
p_j&=p_t\left(\cosh y,\cos\varphi,\sin\varphi,\sinh y\right),\\
k_i&=k_{ti}\left(\cosh\eta_i,\cos\phi_i,\sin\phi_i,\sinh\eta_i\right),
\end{align}
\end{subequations}
where $y$ and $\eta_i$ are the rapidities of the jet and soft emissions, and $\varphi$ and $\phi_i$ are their respective azimuthal angles, measured relative to the beam axis. Neglecting recoil effects is justified at SL accuracy.

We use polar coordinates $(r,\theta)$ in the $\eta-\phi$ plane, centered at the jet-initiating parton $(y,\varphi)$. Each particle $k_i$ is characterized by $r_i$ and $\theta_i$
\begin{subequations}
\begin{align}
\eta_i-y=R\,r_i\cos\theta_i\,,\\
\phi_i-\varphi=R\,r_i\sin\theta_i\,,
\end{align}
\end{subequations}
where $r_i$ is normalized to the jet radius $R$, so that $r_i = 1$ corresponds to the jet boundary.

At Born level, the two relevant partonic subprocesses for Z boson and jet production in $pp$ collisions are $(\delta_g)$: $q\bar{q} \to Zg$ and $(\delta_q)$: $qg \to Zq$. The subscripts $q$ (for quarks/anti-quarks) and $g$ (for gluons) indicate the nature of the outgoing parton initiating the jet.

\subsection{Jet mass, clustering algorithms, and grooming}

We study the normalized invariant jet mass (squared), defined as:
\begin{equation}
\rho=\frac{1}{p_t^2}\left(p_j+\sum_{i\in\mathrm{jet}}k_i\right)^2,
\end{equation}
where the sum includes all soft gluons within the jet. In the massless-quarks limit and under the strong ordering approximation (relevant for NLL resummation), this simplifies to
\begin{align}
\rho&=\sum_{i\in\mathrm{jet}}\frac{2\,k_{ti}}{p_t}\left[\cosh(\eta_i-y)-\cos(\phi_i-\varphi)\right]\notag\\
&\approx\sum_{i\in\mathrm{jet}}\frac{k_{ti}}{p_t}\,R^2\,r_i^2\,,\label{eq:def}
\end{align}
with the final approximation valid at NLL accuracy.

Sequential recombination algorithms define two types of distances: between particle pairs $d_{ij}$ and between particles and the beam $d_{iB}$
\begin{subequations}
\begin{align}
d_{ij} &=\min(k_{ti}^{2p},k_{tj}^{2p})\,R_{ij}^2\,,\quad \text{with } R_{ij}^2=(\eta_i-\eta_j)^2+(\phi_i-\phi_j)^2\,,\\
d_{iB} &=k_{ti}^{2p}\,R^2\,,
\end{align}
\end{subequations}
where $R$ is the jet radius and $p$ determines the clustering algorithm: $p=-1$ for anti-$k_t$ \cite{Cacciari:2008gp}, $p=+1$ for $k_t$ \cite{Catani:1993hr, Ellis:1993tq}, and $p=0$ for Cambridge-Aachen (CA) \cite{Wobisch:1998wt}.

Several jet grooming techniques have been developed to improve jet substructure studies, including mass-drop/filtering \cite{Butterworth:2008iy}, trimming \cite{Krohn:2009th}, and pruning \cite{Ellis:2009su, Ellis:2009me}. In this work, we focus on trimming as described in ref. \cite{Krohn:2009th}. In summary, jets are clustered using a sequential recombination algorithm with a jet radius $R$. We use the anti-$k_t$ algorithm with $R = 0.7$. The jet constituents are then re-clustered into sub-jets with a smaller radius $R_{\mathrm{sub}}$. For this phase, we consider the $k_t$ algorithm with $R_{\mathrm{sub}}=0.2$. Sub-jets with momentum fractions below a chosen threshold $z_{\mathrm{cut}}$ are removed. We set $z_{\mathrm{cut}} = 0.03$. The final trimmed jet comprises all particles from the original jet, excluding those in rejected sub-jets. This trimming procedure is actually less aggressive than pruning but removes more particles than filtering.

\subsection{Resummed trimmed jet mass distribution and global form factor}

The integrated jet mass distribution is formulated as a sum over channels $\delta$, involving a convolution of the differential cross-section $\d\sigma_{0\delta}/\d\mathcal{B}_\delta$ for the Born configuration $\mathcal{B}_\delta$ with the resummed global form factor $f_{\mathcal{B},\delta}(\rho)$, as well as the non-global and clustering functions $\mathcal{S}_\delta(\rho)$ and $\mathcal{C}_\delta(\rho)$ \cite{Banfi:2004yd}
\begin{equation}\label{eq:master}
\sigma(\rho)=\sum_\delta \int \d\mathcal{B}_\delta\,\frac{\d\sigma_{0\delta}}{\d\mathcal{B}_\delta}\,\Omega_\mathcal{B}\,f_{\mathcal{B},\delta}(\rho)\,\mathcal{S}_\delta(\rho)\,\mathcal{C}_\delta(\rho)
\left(1+\alpha_s\,C_1(\mathcal{B}_\delta)+\mathcal{O}(\alpha_s^2)\right),
\end{equation}
where $\Omega_\mathcal{B}$ represents the experimental cuts and $C_1(\mathcal{B}_\delta)$ is the $\mathcal{O}(\alpha_s)$ correction term, introduced to improve the accuracy up to NLL'.  A detailed discussion of this correction term will be provided later.

The global form factor is expressed in the form of an exponential function as follows
\begin{equation}\label{eq:master2}
f_{\mathcal{B},\delta}(\rho) = \frac{1}{\Gamma[1 + \mathcal{R}_\delta'(\rho)]} \exp\left[-\mathcal{R}_\delta(\rho) - \gamma_E\,\mathcal{R}_\delta'(\rho)\right],
\end{equation}
where $\mathcal{R}_\delta$ is the radiator associated with  channel $\delta$, $\gamma_E$ is the Euler-Mascheroni constant, and $\mathcal{R}_\delta'(\rho)$ is the derivative of the radiator with respect to the resummed logarithms.

In the following section, we shall first compute the leading-order expansion in $\alpha_s$ of the form factor $f_{\mathcal{B},\delta}(\rho)$. The full expression of the radiator to NLL accuracy is provided in Appendix \ref{sec:Rad}. We shall then, in Section 4, derive the non-global and clustering corrections at second order in the coupling, and estimate their resummed result at all orders in the large-$\Nc$ approximation. The convolution in eq. \eqref{eq:master} is then performed in Section 5. For now, we begin by evaluating $f^{(1)}_{\mathcal{B},\delta}(\rho)$, the leading-order expansion of the global form factor.

\section{The jet mass distribution at leading order with trimming}

\subsection{Review of the ungroomed distribution}

We begin by reviewing the ungroomed form factor at leading order, as computed in refs. \cite{Dasgupta:2012hg,Ziani:2021dxr}. At first order in the strong coupling, the final state consists of a single soft gluon emission $k_1$ alongside the hard parton initiating the jet $p_j$. Within the strongly ordered regime, the gluon is clustered to the hard jet if its distance from the hard parton in the $\eta-\phi$ plane is actually less than $R$, $R_{1j}<R$. This leads to a non-zero jet mass for real emissions, while the jet mass remains zero in virtual corrections, resulting in a mis-cancellation of real and virtual soft-collinear singularities.

The form factor for channel $\delta$, in the eikonal approximation (sufficient to capture the soft singularities), is given at $\mathcal{O}(\alpha_s)$ for a $k_t$, anti-$k_t$, or CA-clustered jet of radius $R$ by \cite{Dasgupta:2012hg, Ziani:2021dxr}
\begin{equation}\label{eq:f1}
f^{\mathrm{ungr},(1)}_{\mathcal{B},\delta}(\rho)=-\sum_{(i\ell)}\mathcal{C}_{i\ell}\,\frac{R^2}{2}\int\frac{\d\xi_1}{\xi_1}\,\d r_1^2\,\frac{\d\theta_1}{2\pi}\,\bar{\alpha}_{s}\,\omega_{i\ell}^1\, \Theta(\xi_1\,R^2\,r_1^2-\rho)\,\Xi^{\mathrm{ungr}}_{\mathrm{in}}(k_1)\,,
\end{equation}
where the sum is over the dipoles formed by the hard/eikonal legs $(i\ell)=\{(ab),(aj),(bj)\}$. The color factors, $\mathcal{C}_{i\ell}$, depend on the nature of the partons involved: $\mathcal{C}_{i\ell}=\CA=\Nc$ for dipoles involving gluons and $\mathcal{C}_{i\ell}=2\,\CF-\CA=-1/\Nc$ for dipoles involving only quarks or anti-quarks. Here, $\CA=\Nc$ and $\CF=(\Nc^2-1)/(2\Nc)$ are the QCD color factors with $\Nc=3$. The quantity $\xi_1=k_{t1}/p_t$ is the momentum fraction of the soft emission relative to the jet’s transverse momentum $p_t$. The angular antenna function for the emission of the soft gluon $k_1$ from the dipole $(i\ell)$ is given by
\begin{equation}
\omega_{i\ell}^1=\frac{k_{t1}^2}{2}\,\frac{p_i\cdot p_\ell}{(p_i\cdot k_1)(k_1\cdot p_\ell)}\,.
\end{equation}
For the dipole $(ab)$, $\omega_{ab}^1=1$, and for dipoles $(aj)$ and $(bj)$, we have
\begin{subequations}
\begin{align}
\omega_{aj}^1&=\frac{1}{2}\,\frac{\exp(R\,r_1\cos\theta_1)}{\cosh(R\,r_1\cos\theta_1)-\cos(R\,r_1\sin\theta_1)}\,,\\
\omega_{bj}^1&=\frac{1}{2}\,\frac{\exp(-R\,r_1\cos\theta_1)}{\cosh(R\,r_1\cos\theta_1)-\cos(R\,r_1\sin\theta_1)}\,.
\end{align}
\end{subequations}

In eq. \eqref{eq:f1}, $\bar{\alpha}_s=\alpha_s/\pi$ is the ``reduced'' strong coupling. While its argument is irrelevant at leading order, for full resummation it is evaluated at the invariant transverse momentum of the emission $k_1$ in the emitting dipole's rest frame, $\kappa_{t1,(i\ell)}^2=k_{t1}^2/\omega_{i\ell}^1$. For dipole $(ab)$, $\kappa_{t1,(ab)}^2=k_{t1}^2 = \xi_1^2\,p_t^2$, and for dipoles $(aj)$ and $(bj)$ at NLL accuracy, $\kappa_{t1,(aj)}^2=k_{t1}^2 R^2 r_1^2=\xi_1^2\,p_t^2\,R^2\,r_1^2$.

The step function $\Theta(\xi_1\,R^2\,r_1^2-\rho)$ ensures that in the integrated distribution, the jet mass (equal to $\xi_1\,R^2\,r_1^2$ at NLL accuracy) is less than $\rho$ for real emissions. Only virtual emissions with $\xi_1\,R^2\,r_1^2>\rho$ remain uncancelled and are integrated over. Lastly, $\Xi_{\mathrm{in}}(k_1)$ restricts the integration region so that the gluon remains within the jet after clustering. For $k_t$, anti-$k_t$, and CA algorithms, it is simply given by 
\begin{equation}\label{eq:xi1ungr}
\Xi^{\mathrm{ungr}}_{\mathrm{in}}(k_1)=\Theta(R^2-R_{1j}^2) = \Theta(1-r_1^2)\,,
\end{equation}
where $R_{1j}^2=R^2\,r_1^2$.

After performing the integrations, and including corrections due to hard-collinear emissions to the outgoing jet, the NLL result is
\begin{equation}\label{eq:resungr}
f^{\mathrm{ungr},(1)}_{\mathcal{B},\delta}(\rho)=-\left(C_j\,B_j+\sum_{(i\ell)}\mathcal{C}_{i\ell}\,\mathcal{J}_{i\ell}(R^2)\right)\bar{\alpha}_s\ln\frac{R^2}{\rho}-\frac{C_j}{2}\,\bar{\alpha}_s\ln^2\frac{R^2}{\rho}\,,
\end{equation}
where $C_j=(\mathcal{C}_{aj}+\mathcal{C}_{bj})/2$, with $C_j=\CF$ for quark-initiated jets and $C_j=\CA$ for gluon-initiated jets. The $B_j$ term accounts for hard-collinear corrections, and we have
\begin{equation}
\begin{aligned}
B_q &= -\frac{3}{4}                                           &&\text{quark-initiated jets}\,, \\
B_g &= -\frac{11\,\CA-4\,\mathrm{T_R}\,\mathrm{n_f}}{12\,\CA}=-\frac{\pi\beta_0}{\CA} \quad&&\text{gluons-initiated jets}\,,
\end{aligned}
\end{equation}
where $\mathrm{T_R}=1/2$ is the normalization constant for $\mathrm{SU(\Nc)}$, $n_f=5$ is the number of active flavors, and $\beta_0$ is the one-loop coefficient of the QCD beta function. This correction arises from using the full splitting functions in the double logarithmic integration over energy fractions, where $1/\xi_1$ is replaced by $P_j(\xi_1)/\xi_1$ in the integrand. For a quark jet ($j=q$) and a gluon jet ($j=g$), the splitting functions, stripped of color factors, are given by \cite{Banfi:2004yd}
\begin{subequations}
\begin{align}
P_q(\xi)&=\xi\,P_{gq}(\xi)\,,\\
P_g(\xi)&=\xi\,\frac{1}{2}\,P_{gg}(\xi)+\frac{\mathrm{T_R}\,\mathrm{n_f}}{\CA}\,\xi\,P_{qg}(\xi)\,,
\end{align}
\end{subequations}
where
\begin{subequations}
\begin{align}
P_{gq}(\xi)&=\frac{1+(1-\xi)^2}{2\,\xi}\,,\\
P_{gg}(\xi)&=\frac{2(1-\xi)}{\xi}+\xi(1-\xi)\,,\\
P_{qg}(\xi)&=\frac{1}{2}\left[\xi^2+(1-\xi)^2\right].
\end{align}
\end{subequations}

The jet functions $\mathcal{J}_{i\ell}$ are
\begin{subequations}
\begin{align}
\mathcal{J}_{ab}(R^2)&=\frac{R^2}{2}\,,\label{eq:jfab}\\
\mathcal{J}_{aj}(R^2)=\mathcal{J}_{bj}(R^2)&= \frac{R^2}{8}+\frac{R^4}{576}+\frac{R^8}{4\,147\,200}+\mathcal{O}(R^{12})\,.\label{eq:jfaj}
\end{align}
\end{subequations}
These functions result from the angular integration over the dipole antenna functions and are expressed as a truncated series in the jet radius $R$. The truncation is sufficiently accurate for values of $R$ up to order 1.

\subsection{Trimming correction at leading order}

Now, let us examine how trimming affects the jet mass distribution. When the soft gluon $k_1$ is emitted within a distance $R_{\mathrm{sub}}$ from the hard parton in the $\eta-\phi$ plane, it remains in the jet after trimming. Additionally, if this gluon is emitted within an annulus defined by inner radius $R_{\mathrm{sub}}$ and outer radius $R$, and its transverse momentum satisfies $k_{t1}>z_{\mathrm{cut}}\, p_t$, it forms a sub-jet that is retained alongside the hard sub-jet of the Born parton, contributing to the final jet mass. However, if $k_{t1}<z_{\mathrm{cut}}\,p_t$, the sub-jet formed by the soft emission is trimmed away, yielding a zero jet mass. Furthermore, if the soft gluon is emitted at a distance greater than $R$ from the hard parton, it does not get clustered into the jet, also resulting in a vanishing jet mass.

The leading-order form factor $f^{\mathrm{trim},(1)}_{\mathcal{B},\delta}$ for trimmed jets is similar to the ungroomed jet mass form factor in eq. \eqref{eq:f1}, with the replacement of the angular phase space $\Xi^{\mathrm{ungr}}_{\mathrm{in}}(k_1)\to \Xi^{\mathrm{trim}}_{\mathrm{in}}(k_1)$, ensuring that the gluon $k_1$ remains in the jet after trimming. The trimmed angular phase-space is given by
\begin{equation}\label{eq:xi12}
\Xi^{\mathrm{trim}}_{\mathrm{in}}(k_1)=\Theta(R_{\mathrm{sub}}^2-R^2_{1j})+\Theta(R^2_{1j}-R_{\mathrm{sub}}^2)\,\Theta(R^2-R^2_{1j})\,\Theta(\xi_1-z_{\mathrm{cut}})\,.
\end{equation}
Multiplying by the step function $\Theta(\xi_1\,R^2\,r_1^2-\rho) = \Theta(\xi_1-\rho/R_{1j}^2)$, we express the phase space in a form that clearly defines the lower limit on the $\xi_1$ integration, depending on whether $z_{\mathrm{cut}}$ or $\rho/R_{1j}^2$ is smaller. We can then write
\begin{align}
\Theta(\xi_1\,R_{1j}^2-\rho)\,\Xi^{\mathrm{trim}}_{\mathrm{in}}&=\Theta(\xi_1\,R_{1j}^2-\rho)\,\Theta(R_{\mathrm{sub}}^2-R^2_{1j})\notag\\
&+\Theta(\xi_1\,R_{1j}^2-\rho)\,\Theta(R^2_{1j}-R_{\mathrm{sub}}^2)\,\Theta(R^2-R^2_{1j})\,\Theta(\rho/z_{\mathrm{cut}}-R_{1j}^2)\notag\\
&+\Theta(\xi_1-z_{\mathrm{cut}})\,\Theta(R^2_{1j}-R_{\mathrm{sub}}^2)\,\Theta(R^2-R^2_{1j})\,\Theta(R_{1j}^2-\rho/z_{\mathrm{cut}})\,.\label{eq:phs}
\end{align}
The second term in eq. \eqref{eq:phs} defines the integration region for the variable $r_1$ such that
\begin{equation}
R_{\mathrm{sub}}^2<R^2\,r_1^2<\min(R^2,\rho/z_{\mathrm{cut}})\,,
\end{equation}
which contributes only if $\rho>R_{\mathrm{sub}}^2z_{\mathrm{cut}}$. Additionally, the third term in eq. \eqref{eq:phs} yields
\begin{equation}
\max(R_{\mathrm{sub}}^2,\rho/z_{\mathrm{cut}})<R^2\,r_1^2<R^2,
\end{equation}
which contributes only when $\rho<R^2z_{\mathrm{cut}}$.

We thus identify three distinct regions of the $\rho$ variable with two transition points: a large-$\rho$ region where $\rho>R^2z_{\mathrm{cut}}$, an intermediate region $R_{\mathrm{sub}}^2 z_{\mathrm{cut}}<\rho<R^2 z_{\mathrm{cut}}$, and a small-$\rho$ region where $\rho<R^2_{\mathrm{sub}}z_{\mathrm{cut}}$. For the chosen values of $R$, $R_{\mathrm{sub}}$, and $z_{\mathrm{cut}}$, the transition points are $R^2z_{\mathrm{cut}} = 0.015$ and $R_{\mathrm{sub}}^2z_{\mathrm{cut}}=0.0012$. The phase space can be written in a compact form for each region as follows
\begin{itemize}
\item Large-$\rho$ region, $\rho>R^2z_{\mathrm{cut}}$
\begin{equation}
\Theta(\xi_1\,R_{1j}^2-\rho)\,\Xi^{\mathrm{trim}}_{\mathrm{in}}= \Theta(\xi_1\,R_{1j}^2-\rho)\,\Theta(R^2-R_{1j}^2)\,.
\end{equation}
\item Intermediate-$\rho$ region, $R_{\mathrm{sub}}^2z_{\mathrm{cut}}<\rho<R^2z_{\mathrm{cut}}$
\begin{align}
\Theta(\xi_1\,R_{1j}^2-\rho)\,\Xi^{\mathrm{trim}}_{\mathrm{in}}&=\Theta(\xi_1\,R_{1j}^2-\rho)\,\Theta(\rho/z_{\mathrm{cut}}-R_{1j}^2)\notag\\
&+\Theta(\xi_1-z_{\mathrm{cut}})\,\Theta(R^2-R^2_{1j})\,\Theta(R_{1j}^2-\rho/z_{\mathrm{cut}})\,.
\end{align}
\item Small-$\rho$ region,  $\rho<R_{\mathrm{sub}}^2z_{\mathrm{cut}}$
\begin{align}
\Theta(\xi_1\,R_{1j}^2-\rho)\,\Xi^{\mathrm{trim}}_{\mathrm{in}}&=\Theta(\xi_1\,R_{1j}^2-\rho)\,\Theta(R_{\mathrm{sub}}^2-R^2_{1j})\notag\\
&+\Theta(\xi_1-z_{\mathrm{cut}})\,\Theta(R^2-R^2_{1j})\,\Theta(R^2_{1j}-R_{\mathrm{sub}}^2)\,.
\end{align}
\end{itemize}
In the large-$\rho$ region, the angular phase space for the trimmed jet mass form factor is identical to the ungroomed case. This implies that the distribution in this region, at NLL accuracy, is unaffected by trimming and follows the same form as eq.  \eqref{eq:resungr}.

To integrate eq. \eqref{eq:f1} using the phase space in eq. \eqref{eq:xi12}, we expand the antenna functions as a power series in $R$. The result splits into collinear and wide-angle contributions as follows
\begin{equation}
f^{\mathrm{trim},(1)}_{\mathcal{B},\delta}(\rho) = f^{\mathrm{trim},(1)}_{\mathcal{B},\delta,\text{wide}}(\rho) + f^{\mathrm{trim},(1)}_{\mathcal{B},\delta,\text{coll.}}(\rho)\,,
\end{equation}
with the collinear contribution containing double (soft collinear) logarithms and single (hard collinear) logarithms
\begin{align}
f^{\mathrm{trim},(1)}_{\mathcal{B},\delta,\text{coll.}}(\rho)&=-\frac{C_j}{2}\,\bar{\alpha}_s\ln^2\min\left[\frac{R^2}{\rho},\max\left(\frac{1}{z_{\mathrm{cut}}},\frac{R_{\mathrm{sub}}^2}{\rho}\right)\right]
-C_j\,B_j\,\bar{\alpha}_s\ln\frac{R^2}{\rho}\notag\\
&-\Theta(R^2z_{\mathrm{cut}}-\rho)\,C_j\,\bar{\alpha}_s\ln\frac{1}{z_{\mathrm{cut}}}\left[\ln\frac{R^2}{\rho}-\ln\max\left(\frac{1}{z_{\mathrm{cut}}},\frac{R_{\mathrm{sub}}^2}{\rho}\right)\right],\label{eq:f1col}
\end{align}
and the soft wide-angle contribution containing only single logarithms
\begin{align}
f^{\mathrm{trim},(1)}_{\mathcal{B},\delta,\text{wide}}(\rho)&=-\sum_{(i\ell)}\mathcal{C}_{i\ell}\,\mathcal{J}_{i\ell}\left(\min\left[R^2,\max\{\rho/z_{\mathrm{cut}},R_{\mathrm{sub}}^2\}\right]\right)\bar{\alpha}_s\ln\frac{R^2}{\rho}\notag\\
&-\Theta(R^2 z_{\mathrm{cut}}-\rho)\sum_{(i\ell)}\mathcal{C}_{i\ell}\left[\mathcal{J}_{i\ell}(R^2)-\mathcal{J}_{i\ell}(\max[\rho/z_{\mathrm{cut}},R_{\mathrm{sub}}^2])\right]\bar{\alpha}_s\ln\frac{1}{z_{\mathrm{cut}}}\,.
\end{align}

It is important to note that while the $\mathcal{O}(\rho/z_\mathrm{cut})$ terms in the jet functions for the soft wide-angle contribution are formally sub-leading, they have a significant numerical impact in the region where $\rho \sim z_\mathrm{cut}$ and therefore cannot be neglected. 
\footnote{At the transition points between the three regions, we have $\rho/z_{\mathrm{cut}} = R_{\mathrm{sub}}^2$ and $R^2$.} These terms are also necessary to maintain the continuity of the distribution across the transition points. Our finding is consistent with the results obtained in ref. \cite{Dasgupta:2013ihk} for the trimmed jet mass distribution in $e^+e^-$ annihilation to jets.

\subsection{Comparison to Monte Carlo results}

\begin{center}
\begin{figure}[ht]
\centering
\includegraphics[width=0.48\textwidth]{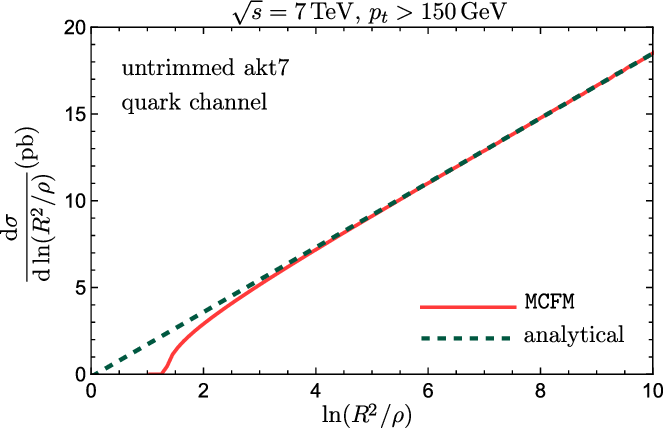}
\includegraphics[width=0.48\textwidth]{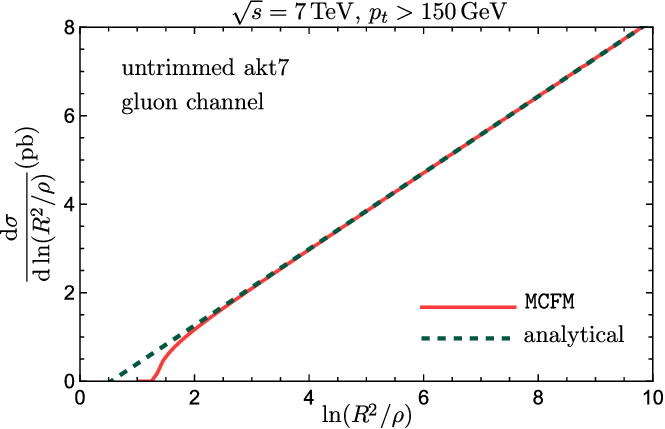}
\includegraphics[width=0.48\textwidth]{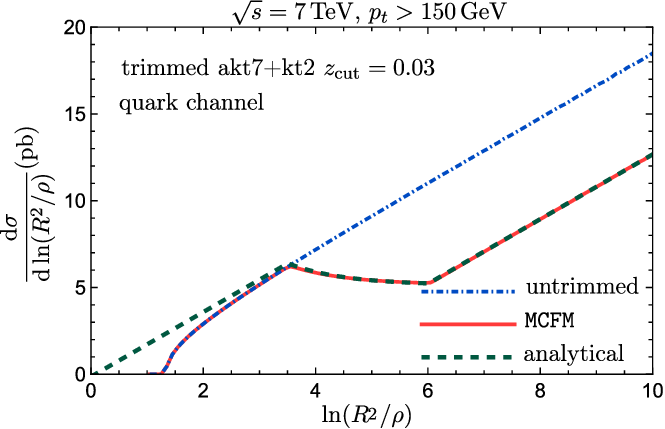}
\includegraphics[width=0.48\textwidth]{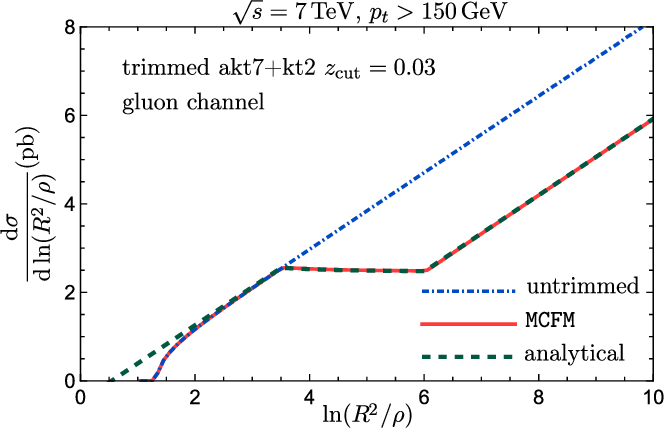}
\caption{\label{fig:00}
Comparison of the differential jet mass distribution $\d\sigma/\d L$ between the analytical result and the Monte Carlo result obtained with \texttt{MCFM}, for both untrimmed and trimmed jets. Separate contributions from quark and gluon channels are shown.}
\end{figure}
\end{center}

To validate our analytical results, we conduct a phenomenological study to compare them with Monte Carlo simulations. Figure \ref{fig:00} displays the analytical differential distribution $\d\sigma/\d L$, obtained by differentiating $\sigma(\rho)$ (using eq. \eqref{eq:master} at $\mathcal{O}(\alpha_s)$) with respect to $L=\ln(R^2/\rho)$. This analytical distribution is compared to the Monte Carlo results obtained using the fixed-order program \texttt{MCFM 9.1}. We also utilized \texttt{MadGraph5\_aMC@NLO}, which produced similar results. Our obtained results are computed using the \texttt{NNPDF30\_nlo} parton distribution functions with $\alpha_s(M_Z)=0.118$, interfaced via \texttt{LHAPDF} \cite{Buckley:2014ana}. The center-of-mass energy is set to $\sqrt{s}=7\,\mathrm{TeV}$, and jets with transverse momentum $p_t>150\,\mathrm{GeV}$ and rapidity $|y|<2.5$ are selected. We set the Z boson decay width to zero to simplify the analysis. The upper plots in figure \ref{fig:00} show the ungroomed distribution for quark (left) and gluon (right) channels, while the lower plots display the trimmed distributions. The jet radius is fixed at $R=0.7$, with substructure parameters $R_{\mathrm{sub}}=0.2$ and $z_{\mathrm{cut}}=0.03$. Renormalization and factorization scales are chosen to be $\mu_\mathrm{R}=\mu_\mathrm{F}=150.0\,\mathrm{GeV}$.

In the log-enhanced region, corresponding to small values of $\rho$, the ungroomed distribution exhibits a linear behavior, confirming the expected structure
\begin{equation}
\frac{\d\sigma_{\delta}}{\d L} = \sigma_{0\delta}\left(g_{11,\delta} + 2\,g_{12,\delta}\,L\right),
\end{equation}
where $g_{11}$ and $g_{12}$ are the coefficients of the single and double logarithms, respectively. In this region, we observe excellent agreement between the Monte Carlo and analytical results, consistent with prior studies \cite{Dasgupta:2012hg, Ziani:2021dxr}.

The trimmed distribution also shows strong agreement with the analytical results across the three regions of jet mass. At large $\rho$  (small $L$), the ungroomed and trimmed distributions converge. In the intermediate region, the trimmed distribution deviates from the double logarithmic behavior in $\rho$, with the logarithms of $\rho$  being replaced by those in $z_{\mathrm{cut}}$ (see eq. \eqref{eq:f1col}). The differential distribution $\d\sigma/\d L$ remains approximately constant, with any small variations attributable to the $\rho$-dependence of the jet functions. Finally, in the small-$\rho$ region, the double logarithmic behavior re-emerges with the same coefficient $\sigma_{0\delta}\,\bar{\alpha}_s\,C_j/2$ as in the ungroomed case, while the single logarithm ($y$-intercept) is reduced due to the influence of $R_{\mathrm{sub}}$ and $z_{\mathrm{cut}}$ dependence.

\subsection{NLL' corrections}

To achieve NLL' accuracy in the expansion of the cumulative distribution, we must include all terms of the form $\alpha_s^n\,L^{2n-2}$ at $n^{\mathrm{th}}$ order. These terms can be captured by calculating the fixed-order constant $C_1^{(\delta)}$, averaged over the Born configuration $\mathcal{B}_\delta$, as defined in ref. \cite{Banfi:2010xy}
\begin{equation}
\alpha_s C_1^{(\delta)}\equiv\left\langle\alpha_s C_1(\mathcal{B}_\delta)\right\rangle = \frac{1}{\sigma_{0\delta}}\lim_{\rho \to 0}\left[\int_0^\rho\frac{\d\sigma_{\mathrm{NLO}}^{(\delta)}}{\d \rho'}\d \rho'-\sigma_{\mathrm{NLL},\alpha_s}^{(\delta)}(\rho)\right],
\end{equation}
where $\d\sigma_{\mathrm{NLO}}^{(\delta)}/\d \rho$ represents the differential distribution obtained from \texttt{MCFM}, and $\sigma_{\mathrm{NLL},\alpha_s}^{(\delta)}(\rho)$ is the NLL-resummed distribution expanded to $\mathcal{O}(\alpha_s)$. 
The computed values of the constant $\alpha_sC_1^{(\delta)}$, along with the Born and NLO cross-sections, are summarized in table \ref{tab:C1}.

\begin{table}[ht]
\centering
\begin{tabular}{|c|c|c|c|c|c|c|}
\hline
           & $\sigma_{0g}$(pb) & $\sigma_{0q}$(pb) & $\sigma_{1g}$(pb) & $\sigma_{1q}$(pb) & $\alpha_sC_{1}^{(g)}$ & $\alpha_sC_{1}^{(q)}$ \\ \hline
ungroomed  & 8.23         & 40.0        & 14.6        & 31.8        & 1.73    & 0.722    \\ \hline
trimmed    & 8.23         & 40.0        & 14.4        & 31.3        & 1.71    & 0.733    \\ \hline
\end{tabular}
\caption{\label{tab:C1}Numerical results for the Born cross-section, NLO cross-section, and the constant $\alpha_sC_{1}^{(\delta)}$.}
\end{table}

\section{Two-loop calculation: NGLs and CLs}

\subsection[Ungroomed NGLs with anti-\texorpdfstring{$k_t$}{kt} and \texorpdfstring{$k_t$}{kt} clustering]{Ungroomed NGLs with anti-$\boldsymbol{k_t}$ and $\boldsymbol{k_t}$ clustering}

We begin by reviewing and extending the results for the ungroomed NGLs. At $\mathcal{O}(\alpha_s^2)$, NGLs arise from the correlated emission of two strongly-ordered gluons $k_1$ and $k_2$, where $k_1$ lies outside the jet and $k_2$ inside. The non-global contribution at this order is given by \cite{Dasgupta:2012hg,Ziani:2021dxr}
\begin{subequations}
\begin{align}
f^{\mathrm{ungr},(2),\mathrm{NG}}_{\mathcal{B},\delta}(\rho)&=-\frac{1}{2}\,\bar{\alpha}_s^2\ln^2\frac{R^2}{\rho}\,\CA \sum_{(i\ell)}\mathcal{C}_{i\ell}\,\mathcal{G}^{\mathrm{ungr}}_{2,(i\ell)}(R^2)\,,\\
\mathcal{G}^{\mathrm{ungr}}_{2,(i\ell)}(R^2) &= \frac{R^4}{4}\int \d r_1^2\,\frac{\d\theta_1}{2\pi}\int \d r_2^2\,\frac{\d\theta_2}{2\pi}\, \Xi^{\mathrm{ungr}}_{\mathrm{NG}}(k_1,k_2)\,\mathcal{A}_{i\ell}^{12}\,,
\end{align}
\end{subequations}
where the two-loop non-Abelian antenna function is defined as
\begin{equation}
\mathcal{A}_{i\ell}^{12} =\omega_{i\ell}^1\left(\omega_{i1}^2+\omega_{1\ell}^2-\omega_{i\ell}^2\right).
\end{equation}
The phase-space constraint for anti-$k_t$ clustering is given by \footnote{Note that the edge of phase space restricts $R\, r_i|\sin\theta_i|<\pi$.}
\begin{equation}
\Xi^{\mathrm{ungr,ak_t}}_{\mathrm{NG}}(k_1,k_2) =\Theta(R_{1j}^2-R^2)\,\Theta(R^2-R_{2j}^2)\,,
\end{equation}
while for $k_t$ clustering, it takes the form
\begin{equation}\label{eq:pskt}
\Xi^{\mathrm{ungr,k_t}}_{\mathrm{NG}}(k_1,k_2) =\Theta(R_{1j}^2-R^2)\Theta(R^2-R_{2j}^2)\,\Theta(R_{12}^2-R_{2j}^2)\,,
\end{equation}
where the condition $R_{12}^2 > R_{2j}^2$ ensures that gluon $k_1$ does not pull $k_2$ out of the jet.

To facilitate the calculation of the NGLs coefficient for the trimmed distribution, we introduce modified phase-space functions
\begin{align}
\widetilde{\Xi}_{\mathrm{NG}}^{\mathrm{ak_t}}(R^2,R_{\mathrm{sub}}^2)&=\Theta(R_{1j}^2-R^2)\,\Theta(R_{\mathrm{sub}}^2-R_{2j}^2)\,,\\
\widetilde{\Xi}_{\mathrm{NG}}^{\mathrm{k_t}}(R^2,R_{\mathrm{sub}}^2)&=\Theta(R_{1j}^2-R^2)\,\Theta(R_{\mathrm{sub}}^2-R_{2j}^2)\,\Theta(R_{12}^2-R_{2j}^2)\notag\\
&=\widetilde{\Xi}_{\mathrm{NG}}^{\mathrm{ak_t}}(R^2,R_{\mathrm{sub}}^2)-\Theta(R_{1j}^2-R^2)\Theta(R_{\mathrm{sub}}^2-R_{2j}^2)\Theta(R_{2j}^2-R_{12}^2)\,,\label{eq:sub}
\end{align}
where, for $R_{\mathrm{sub}}<R/2$, the step function $\Theta(R_{2j}^2-R_{12}^2)$ vanishes, reducing $\widetilde{\Xi}_{\mathrm{NG}}^{\mathrm{k_t}}(R^2, R_{\mathrm{sub}}^2)$ to $\widetilde{\Xi}_{\mathrm{NG}}^{\mathrm{ak_t}}(R^2, R_{\mathrm{sub}}^2)$. These phase spaces match the anti-$k_t$ and $k_t$ ungroomed cases when $R_{\mathrm{sub}} = R$.

We then define the non-global factor for dipole $(i\ell)$ as
\begin{align}
\mathcal{G}_{2,(i\ell)}(R^2,R^2_{\mathrm{sub}})&=\frac{R^4}{4}\int \d r_1^2\,\frac{\d\theta_1}{2\pi}\int \d r_2^2\,\frac{\d\theta_2}{2\pi}\,\widetilde{\Xi}_{\mathrm{NG}}(R^2,R_{\mathrm{sub}}^2)\,\mathcal{A}_{i\ell}^{12}\,,
\end{align}
such that $\mathcal{G}_{2,(i\ell)}^{\mathrm{ungr,ak_t}}(R^2)=\mathcal{G}^{\mathrm{ak_t}}_{2,(i\ell)}(R^2,R^2)$ and $\mathcal{G}_{2,(i\ell)}^{\mathrm{ungr,k_t}}(R^2)=\mathcal{G}^{\mathrm{k_t}}_{2,(i\ell)}(R^2,R^2)$. Expanding the integrand as a power series in $R$ and performing the relevant integrations we obtain the following results
\begin{subequations}
\begin{align}
\mathcal{G}^{\mathrm{ak_t}}_{2,(ab)}(R^2,R^2_{\mathrm{sub}})&=-\frac{1}{2}\,R^2\ln R^2+\left(\mathcal{J}_{ab}(R^2)-\mathcal{J}_{ab}(R_{\mathrm{sub}}^2)\right)\ln\left(R^2-R_{\mathrm{sub}}^2\right)\notag\\
&+\frac{1}{2}\,R_{\mathrm{sub}}^2 +\frac{1}{8}\,R^2R_{\mathrm{sub}}^2-\frac{1}{576}\,R^4 R_{\mathrm{sub}}^2-\frac{1}{576}\,R^2R_{\mathrm{sub}}^4+\mathcal{O}(R^8)\,,\\
\mathcal{G}^{\mathrm{ak_t}}_{2,(aj)}(R^2,R^2_{\mathrm{sub}})&=\mathcal{G}^{\mathrm{ak_t}}_{2,(bj)}(R^2,R^2_{\mathrm{sub}})=\left(\mathcal{J}_{aj}(R^2)-\mathcal{J}_{aj}(R_{\mathrm{sub}}^2)\right)\ln\frac{R^2-R_{\mathrm{sub}}^2}{R^2}\notag\\
&+\frac{1}{2}\,\mathrm{Li}_2\frac{R_{\mathrm{sub}}^2}{R^2}+\frac{5}{576}\,R^2R_{\mathrm{sub}}^2-\frac{1}{192}\,R_{\mathrm{sub}}^4+\mathcal{O}(R^8)\,.
\end{align}
\end{subequations}
for anti-$k_t$ clustering, and
\begin{subequations}\label{eq:ser}
\begin{align}
\mathcal{G}^{\mathrm{k_t}}_{2,(ab)}(R^2,R_{\mathrm{sub}}^2)&=\mathcal{G}^{\mathrm{ak_t}}_{2,(ab)}(R^2,R_{\mathrm{sub}}^2)\notag\\
&-\Theta(2\,R_{\mathrm{sub}}-R)\left[\chi_{2}(x)\,R^2+\chi_{4}(x)\,R^4+\chi_{6}(x)\,R^6+\mathcal{O}(R^{8})\right],\\
\mathcal{G}^{\mathrm{k_t}}_{2,(aj)}(R^2,R_{\mathrm{sub}}^2)&=\mathcal{G}^{\mathrm{k_t}}_{2,(bj)}(R^2,R_{\mathrm{sub}}^2)=\mathcal{G}^{\mathrm{ak_t}}_{2,(aj)}(R^2,R_{\mathrm{sub}}^2)\notag\\
&-\Theta(2\,R_{\mathrm{sub}}-R)\left[\psi_{0}(x)+\psi_{2}(x)\,R^2+\psi_{4}(x)\,R^4+\psi_{6}(x)\,R^6+\mathcal{O}(R^{8})\right],
\end{align}
\end{subequations}
for $k_t$ clustering, where the higher-order terms, $\mathcal{O}(R^8)$, are negligibly small.

The coefficient functions $\chi_i(x)$ (for the dipole $(ab)$) and $\psi_i(x)$ (for the dipoles $(aj)$ and $(bj)$), which appear in the $k_t$ clustering non-global factors, depend on the ratio $x = R_{\mathrm{sub}} / R$, with $x > 1/2$. These coefficients have been evaluated numerically and are presented in Table \ref{tab:coef} for selected values of $x$.

\begin{table}[ht]
\centering
\begin{tabular}{|c|c|c|c|c|c|c|c|}
  \hline
$x$ & $\chi_2$ & $\chi_4$ & $\chi_6$ & $\psi_0$ & $\psi_2$ & $\psi_4$ & $\psi_6$\\ \hline 
$0.5$ &  0.0000 & $\phantom{-}0.0000$& 0.0000    & 0.0000 & 0.0000 & 0.000 & 0.000 \\ \hline
$0.6$ &  0.0047 & $-0.0003$ & $2.3\times10^{-6}$ & 0.0076 & 0.0011 & $-2.1\times10^{-5}$ & $-1.3\times10^{-6}$ \\ \hline
$0.7$ &  0.0288 & $-0.0019$ & $1.6\times10^{-5}$ & 0.0394 & 0.0062 & $-0.00012$ &  $-9.4\times10^{-6}$ \\ \hline
$0.8$ &  0.0881 & $-0.0060$ & $5.5\times10^{-5}$ & 0.1040 & 0.0183 & $-0.00035$ &  $-3.3\times 10^{-5}$\\ \hline
$0.9$ &  0.2080 & $-0.0135$ & 0.00014            & 0.2167 & 0.0430 & $-0.00063$ & $-8.8\times10^{-5}$\\ \hline
$1.0$ &  0.4846 & $-0.0258$ & 0.00031            & 0.4569 & 0.1034 & $-0.00038$ & $-0.00019$ \\ \hline
\end{tabular}
\caption{The numerically computed coefficients of the series in eq. \eqref{eq:ser} for the $k_t$ clustering non-global factors, evaluated for selected values of $x=R_{\mathrm{sub}}/R$. For $x\leq 1/2$, these coefficients vanish.\label{tab:coef}}
\end{table}

The results for the non-global factors $\mathcal{G}_{2,(i\ell)}(R^2,R_{\mathrm{sub}}^2)$ have been validated through numerical integration and provide a good approximation even for $R$ of order unity. When $R_{\mathrm{sub}} = R$, the results are consistent with those found in refs. \cite{Dasgupta:2012hg,Ziani:2021dxr} for the non-global coefficients in the ungroomed distribution
\begin{subequations}
\begin{align}
\mathcal{G}^{\mathrm{ak_t}}_{2,(ab)}(R^2,R^2)&=-\frac{1}{2}\,R^2\ln R^2+\frac{1}{2}\,R^2 +\frac{1}{8}\,R^4-\frac{1}{288}\,R^6 +\mathcal{O}(R^8)\,,\\
\mathcal{G}^{\mathrm{ak_t}}_{2,(aj)}(R^2,R^2)&=\mathcal{G}_{2,(bj)}(R^2,R^2)=\frac{\pi^2}{12}+\frac{1}{288}\,R^4+\mathcal{O}(R^8)\,.\\
\mathcal{G}^{\mathrm{kt}}_{2,(ab)}(R^2,R^2)&=\mathcal{G}^{\mathrm{ak_t}}_{2,(ab)}(R^2)-0.4846\,R^2+0.0258\,R^4-0.0003\,R^6+\mathcal{O}(R^{8})\,,\\
\mathcal{G}^{\mathrm{kt}}_{2,(aj)}(R^2,R^2)&=\mathcal{G}^{\mathrm{ak_t}}_{2,(aj)}(R^2)-0.4569-0.1034\,R^2+0.0004\,R^4+0.0002\,R^6+\mathcal{O}(R^{8})\,.
\end{align}
\end{subequations}

\subsection{Trimmed NGLs}

Now we calculate the non-global contribution to the trimmed jet mass distribution at two loops. The jet is defined by applying anti-$k_t$ clustering to the final states, followed by the trimming procedure using $k_t$ clustering for the sub-jets.

NGLs arise from a mis-cancellation between real and virtual contributions when the secondary soft emission $k_2$ remains within the groomed jet, while the harder emission $k_1$ is outside the jet. This scenario can occur in two ways: (1) gluon $k_1$ is outside the jet from the beginning, while gluon $k_2$ remains inside and untrimmed, or (2) both gluons $k_1$ and $k_2$ are initially inside the jet, but $k_1$ is trimmed away while $k_2$ remains untrimmed. In the latter case, gluon $k_2$ must be $k_t$-clustered into the jet at the substructure stage, implying $R_{\mathrm{sub}}^2 > R_{2j}^2$ and $R_{12}^2 > R_{2j}^2$.

The corresponding angular phase space is given by the expression
\begin{align}
\Xi^{\mathrm{trim}}_{\mathrm{NG}}(k_1,k_2)&= \Theta(R_{1j}^2-R^2)\,\Theta(R_{\mathrm{sub}}^2-R_{2j}^2)\notag\\
&+\Theta(R_{1j}^2-R^2)\,\Theta(R^2-R_{2j}^2)\,\Theta(R_{2j}^2-R_{\mathrm{sub}}^2)\,\Theta(\xi_2-z_{\mathrm{cut}})\notag\\
&+\Theta(R^2-R_{1j}^2)\,\Theta(R_{1j}^2-R_{\mathrm{sub}}^2)\,\Theta(z_{\mathrm{cut}}-\xi_1)\,\Theta(R_{\mathrm{sub}}^2-R_{2j}^2)\,\Theta(R_{12}^2-R_{2j}^2)\,.
\end{align}
This phase space is further constrained by the following cuts on gluon energies
\begin{equation}
\Phi(\xi_1,\xi_2)=\Theta(\xi_1-\xi_2)\,\Theta(\xi_2\,R_{2j}^2-\rho)\,.
\end{equation}

The phase space can be simplified in terms of the $\rho$ variable regions, found at one loop, as follows
\begin{itemize}
\item Large-$\rho$ region, $\rho> R^2 z_{\mathrm{cut}}$
\begin{align}
\Phi(\xi_1,\xi_2)\,\Xi^{\mathrm{trim}}_{\mathrm{NG}}(k_1,k_2)=\Phi(\xi_1,\xi_2)\,\Theta(R_{1j}^2-R^2)\,\Theta(R^2-R_{2j}^2)\,.
\end{align}
\item Intermediate-$\rho$ region, $R_{\mathrm{sub}}^2z_{\mathrm{cut}}<\rho <R^2 z_{\mathrm{cut}}$
\begin{align}
\Phi(\xi_1,\xi_2)\,\Xi^{\mathrm{trim}}_{\mathrm{NG}}&=\left[\Phi(\xi_1,\xi_2)-\Theta(\xi_1-\xi_2)\,\Theta(\xi_2-z_{\mathrm{cut}})\right]\Theta(R_{1j}^2-R^2)\,\Theta(\rho/z_{\mathrm{cut}}-R_{2j}^2)\notag\\
&+\Theta(\xi_1-\xi_2)\,\Theta(\xi_2-z_{\mathrm{cut}})\,\Theta(R_{1j}^2-R^2)\,\Theta(R^2-R_{2j}^2)\,.
\end{align}
\item Small-$\rho$ region, $\rho<R_{\mathrm{sub}}^2 z_{\mathrm{cut}}$
\begin{align}
\Phi(\xi_1,\xi_2)\,\Xi^{\mathrm{trim}}_{\mathrm{NG}}&= \left[\Phi(\xi_1,\xi_2)-\Theta(\xi_1-\xi_2)\,\Theta(\xi_2-z_{\mathrm{cut}})\right]\Theta(R_{1j}^2-R^2)\,\Theta(R_{\mathrm{sub}}^2-R_{2j}^2)\notag\\
&+\Theta(\xi_1-\xi_2)\,\Theta(\xi_2-z_{\mathrm{cut}})\,\Theta(R_{1j}^2-R^2)\,\Theta(R^2-R_{2j}^2)\notag\\
&+\Phi(\xi_1,\xi_2)\,\Theta(z_{\mathrm{cut}}-\xi_1)\,\Theta(R^2-R_{1j}^2)\,\Theta(R_{1j}^2-R_{\mathrm{sub}}^2)\notag\\
&\times\Theta(R_{\mathrm{sub}}^2-R_{2j}^2)\,\Theta(R_{12}^2-R_{2j}^2)\,\Theta(R_{2j}^2-\rho/z_{\mathrm{cut}})\,.
\end{align}
\end{itemize}
In the large-$\rho$ region, the phase space is identical to that which generates the NGLs for the ungroomed jet mass distribution, $\widetilde{\Xi}_{\mathrm{NG}}^{\mathrm{ak_t}}(R^2, R^2)$. Therefore, trimming has no effect in this region, similar to the one-loop case.

At this order, the result can be expressed as
\begin{align}
f^{\mathrm{trim},(2),\mathrm{NG}}_{\mathcal{B},\delta}(\rho)&=-\frac{1}{2}\,\bar{\alpha}^2_s\,\CA\sum_{(i\ell)}\mathcal{C}_{i\ell}\,\bigg(\ln^2\min\left[\frac{R^2}{\rho},\frac{1}{z_{\mathrm{cut}}}\right]\mathcal{G}^{\mathrm{ak_t}}_{2,(i\ell)}(R^2,R^2)\notag\\
&+\Theta(R^2z_{\mathrm{cut}}-\rho)\left[\ln^2\frac{R^2}{\rho}-\ln^2\frac{1}{z_{\mathrm{cut}}}\right]\mathcal{G}^{\mathrm{ak_t}}_{2,(i\ell)}(R^2, \max[\rho/z_{\mathrm{cut}} ,R_{\mathrm{sub}}^2])\notag\\
&+\Theta(R_{\mathrm{sub}}^2z_{\mathrm{cut}}-\rho)\ln^2\frac{R_{\mathrm{sub}}^2z_{\mathrm{cut}}}{\rho}
\left[\mathcal{G}^{\mathrm{k_t}}_{2,(i\ell)}(R_{\mathrm{sub}}^2,R_{\mathrm{sub}}^2)-\mathcal{G}^\mathrm{k_t}_{2,(i\ell)}(R_{\mathrm{sub}}^2,\rho/z_{\mathrm{cut}})\right.\notag\\
&\left.+\mathcal{G}^{\mathrm{k_t}}_{2,(i\ell)}(R^2,\rho/z_{\mathrm{cut}})-\mathcal{G}^{\mathrm{k_t}}_{2,(i\ell)}(R^2,R_{\mathrm{sub}}^2)\right]\bigg)\,.\label{eq:NGres}
\end{align}

\subsection[Ungroomed CLs with \texorpdfstring{$k_t$}{kt} clustering]{Ungroomed CLs with $\boldsymbol{k_t}$ clustering}

We recall that the ungroomed CLs emerge specifically when the $k_t$ clustering algorithm is applied to the final states rather than the anti-$k_t$ algorithm. At the two loop level, these logarithms arise when two strongly-ordered gluons are emitted directly from the hard partons. In this scenario, the harder gluon $k_1$ lies outside the jet while the softer gluon $k_2$ is initially inside. However, $k_2$ is closer to $k_1$ than to the jet-initiating parton, leading to $k_2$ being pulled out by $k_1$ and consequently leaving the jet massless. In the virtual corrections, when $k_1$ is virtual, the gluon $k_2$ remains inside the jet resulting in a non-zero jet mass. The mismatch between real and virtual contributions gives the clustering logarithmic term
\begin{align}
f^{\mathrm{ungr},(2),\mathrm{CL}}_{\mathcal{B},\delta}(\rho)&=\frac{1}{2}\,\bar{\alpha}_s^2\ln^2\frac{R^2}{\rho}\sum_{(ik,\ell m)}\mathcal{C}_{ik}\,\mathcal{C}_{\ell m}\,\mathcal{F}^{\mathrm{ungr}}_{2,(ik,\ell m)}(R^2)\,,\\
\mathcal{F}^{\mathrm{ungr}}_{2,(ik,\ell m)}(R^2)&=\frac{R^4}{4}\int\d r_1^2\,\frac{\d\theta_1}{2\pi}\int \d r_2^2\,\frac{\d\theta_2}{2\pi}\,\Xi^{\mathrm{ungr}}_{\mathrm{CL}}(k_1,k_2)\,\omega_{ik}^1\,\omega_{\ell m}^2\,,
\end{align}
where the ungroomed phase space is defined as
\begin{equation}
\Xi^{\mathrm{ungr}}_{\mathrm{CL}}(k_1,k_2)=\Theta(R_{1j}^2-R^2)\,\Theta(R^2-R_{2j}^2)\,\Theta(R_{2j}^2-R_{12}^2)\,.
\end{equation}

To introduce the trimmed CLs more effectively, we define the modified phase space
\begin{equation}
\widetilde{\Xi}_{\mathrm{CL}}(R^2,R_{\mathrm{sub}}^2)=\Theta(R_{1j}^2-R^2)\,\Theta(R_{\mathrm{sub}}^2-R_{2j}^2)\,\Theta(R_{2j}^2-R_{12}^2)\,,
\end{equation}
such that $\Xi^{\mathrm{ungr}}_{\mathrm{CL}}(k_1,k_2)=\widetilde{\Xi}_{\mathrm{CL}}(R^2,R^2)$. We define the clustering factor for dipole pairs $(ik,\ell m)$ as
\begin{align}
\mathcal{F}_{2,(ik,\ell m)}(R^2,R_{\mathrm{sub}}^2) &=\frac{R^4}{4}\int\d r_1^2\,\frac{\d\theta_1}{2\pi}\int \d r_2^2\,\frac{\d\theta_2}{2\pi}\,\widetilde{\Xi}_{\mathrm{CL}}(R^2,R_{\mathrm{sub}}^2)\,\omega_{ik}^1\,\omega_{\ell m}^2\,.
\end{align}
Similar to the non-global factor for $k_t$ clustering, these coefficients vanish when the fraction $x=R_{\mathrm{sub}}/R$ is less than $1/2$. The clustering factors can be expanded as a series in the radius parameter $R$, with coefficients dependent on $x$ as follows
\begin{subequations}\label{eq:cff}
\begin{align}
\mathcal{F}_{2,(ab,ab)}&=\gamma(x)\,R^4\,,\\
\mathcal{F}_{2,(ab,aj)}&=\mathcal{F}_{2,(ab,bj)}= \mu_2(x)\,R^2+\mu_4(x)\,R^4+\mu_6(x)\,R^6+\mathcal{O}(R^8)\,,\\
\mathcal{F}_{2,(aj,ab)}&=\mathcal{F}_{2,(bj,ab)}= \sigma_2(x)\,R^2+\sigma_4(x)\,R^4+\sigma_6(x)\,R^6+\mathcal{O}(R^8)\,,\\
\mathcal{F}_{2,(aj,aj)}&=\mathcal{F}_{2,(bj,bj)}= \tau_0(x)+\tau_2(x)\,R^2+\tau_4(x)\,R^4+\tau_6(x)\,R^6+\mathcal{O}(R^8)\,,\\
\mathcal{F}_{2,(aj,bj)}&=\mathcal{F}_{2,(bj,aj)}= \nu_0(x)+\nu_2(x)\,R^2+\nu_4(x)\,R^4+\nu_6(x)\,R^6+\mathcal{O}(R^8)\,.
\end{align}
\end{subequations}
The coefficient functions $\gamma(x)$, $\mu_i(x)$, $\sigma_i(x)$, $\tau_i(x)$, and $\nu_i(x)$, are evaluated analytically when possible, otherwise numerically, as detailed in Appendix \ref{sec:CLs}. 
For $R_{\mathrm{sub}}=R$, these CL coefficients agree with those found in ref. \cite{Ziani:2021dxr}
\begin{subequations}
\begin{align}
&\mathcal{F}_{2,(ab,ab)}(R^2,R^2) = 0.0517\,R^4\,,\\
&\mathcal{F}_{2,(ab,aj)} = \mathcal{F}_{2,(ab,bj)}= 0.0715\,R^2+0.0129\,R^4+0.0003\,R^6+\mathcal{O}(R^8)\,,\\
&\mathcal{F}_{2,(aj,ab)} = \mathcal{F}_{2,(bj,ab)}= 0.0319\,R^2+0.0129\,R^4+0.0006\,R^6+\mathcal{O}(R^8)\,,\\
&\mathcal{F}_{2,(aj,aj)} = \mathcal{F}_{2,(bj,bj)}= 0.0457+0.0475\,R^2+0.0091\,R^4+0.0004\,R^6+\mathcal{O}(R^8)\,,\\
&\mathcal{F}_{2,(aj,bj)} = \mathcal{F}_{2,(bj,aj)}= 0.0457+0.0042\,R^2+0.0004\,R^4+0.00004\,R^6+\mathcal{O}(R^8)\,.
\end{align}
\end{subequations}

\subsection{Trimmed CLs}

When grooming is applied, using the anti-$k_t$ algorithm for both seed-jet finding and sub-jet substructure eliminates CLs contributions. However, as considered in this paper, employing anti-$k_t$ for jet finding followed by $k_t$ clustering for sub-jet substructure introduces a CLs contribution. The phase space in this case is given by
\begin{align}
\Xi_{\mathrm{CL}}^{\mathrm{trim}}(k_1,k_2)&=\Theta(R^2-R_{1j}^2)\,\Theta(R_{1j}^2-R_{\mathrm{sub}}^2)\,\Theta(R_{\mathrm{sub}}^2-R_{2j}^2)\,\Theta(R_{2j}^2-R_{12}^2)\,\Theta(z_{\mathrm{cut}}-\xi_1)\,.
\end{align}
This phase space is further constrained by the cuts $\Phi(\xi_1,\xi_2)=\Theta(\xi_1-\xi_2)\,\Theta(\xi_2\,R_{2j}^2-\rho)$, analogous to the non-global case.

In this setup, $k_1$ lies within the annular region $R_{\mathrm{sub}} < R_{1j} < R$, while $k_2$ is located in the hard sub-jet region, $R_{2j} < R_{\mathrm{sub}}$. Both gluons $k_1$ and $k_2$ are initially clustered to the seed jet via anti-$k_t$. During the substructure phase, $k_t$ clustering is applied to the seed-jet constituents. If $R_{2j} < R_{12}$, $k_2$ evades clustering with the hard jet and instead clusters with $k_1$. The sub-jet formed by $k_1$ and $k_2$ is then rejected if $k_{t1} < z_{\mathrm{cut}}p_t$. On the other hand, if $k_1$ is virtual, $k_2$ remains within the hard sub-jet, leading to a non-zero jet mass.

The phase space can then be expressed as
\begin{align}
\Phi(\xi_1,\xi_2)\,\Xi_{\mathrm{CL}}^{\mathrm{trim}}&=\Phi(\xi_1,\xi_2)\,\Theta(z_{\mathrm{cut}}-\xi_1)\,\Theta(R_{1j}^2-R_{\mathrm{sub}}^2)\notag\\
&\times\Theta(R_{\mathrm{sub}}^2-R_{2j}^2)\,\Theta(R_{2j}^2-R_{12}^2)\,\Theta(R_{2j}^2- \rho/z_{\mathrm{cut}})\,,
\end{align}
assuming $R_{\mathrm{sub}} < R / 2$. This assumption implies that $\Theta(R^2 - R_{1j}^2) = 1$, making the phase space independent of $R$. Consequently, CLs only appear in the small-$\rho$ region, where $\rho < z_{\mathrm{cut}}R^2_{\mathrm{sub}}$. Therefore, the result is written as
\begin{align}
f^{\mathrm{trim},(2),\mathrm{CL}}_{\mathcal{B},\delta}(\rho)&=\Theta(R_{\mathrm{sub}}^2z_{\mathrm{cut}}-\rho)\,\frac{1}{2}\,\bar{\alpha}_s^2\ln^2\frac{R_{\mathrm{sub}}^2z_{\mathrm{cut}}}{\rho} 
\sum_{(ik,\ell m)}\mathcal{C}_{ik}\,\mathcal{C}_{\ell m}\left[\mathcal{F}_{2,(ik,\ell m)}(R_{\mathrm{sub}}^2,R_{\mathrm{sub}}^2)\notag\right.\\
&\left.-\Theta\left(\rho-\frac{1}{4}\,R_{\mathrm{sub}}^2z_{\mathrm{cut}}\right)\mathcal{F}_{2,(ik,\ell m)}(R_{\mathrm{sub}}^2,\rho/z_{\mathrm{cut}})\right].
\end{align}

\begin{figure}
  \centering
  \includegraphics[width=0.49\textwidth]{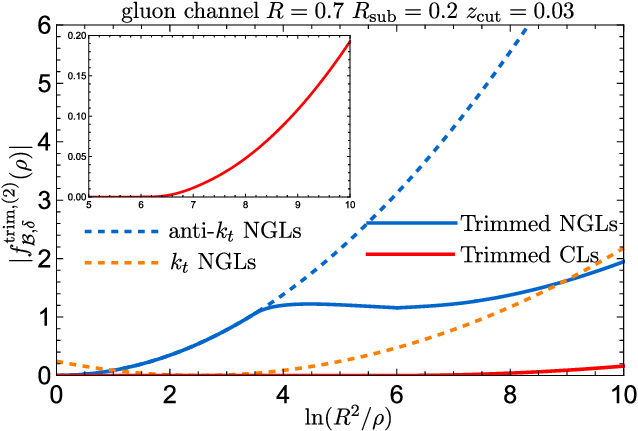}
  \includegraphics[width=0.49\textwidth]{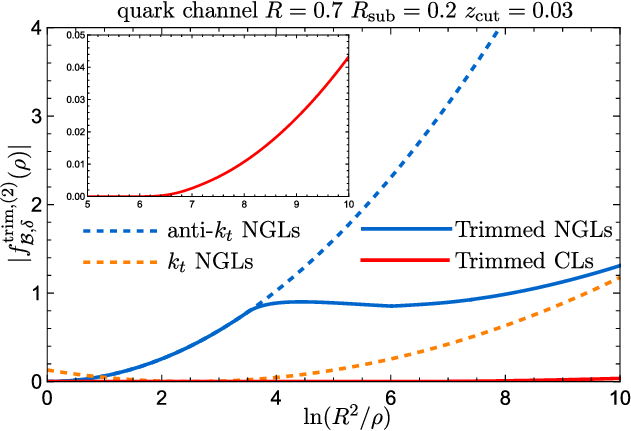}
  \caption{\label{fig:NG}NGLs and CLs contributions to the jet mass distribution at second order. The NGLs contribution is negative while the CLs is positive. The absolute values of the results are shown for comparison.} 
\end{figure}

In the figure \ref{fig:NG}, we present the clustering and non-global contributions, $f^{\mathrm{trim},(2),\mathrm{NG}}_{\mathcal{B},\delta}(\rho)$ and $f^{\mathrm{trim},(2),\mathrm{CL}}_{\mathcal{B},\delta}(\rho)$, for both channels $\delta$. Additionally, we show the NGLs contribution for anti-$k_t$ jets with $R = 0.7$.

We observe that the NGLs contribution is significantly larger with anti-$k_t$ clustering compared to $k_t$ clustering. When applying the trimming algorithm, the NGLs in the large-$\rho$ region coincide with those of anti-$k_t$ clustering but start to deviate in the intermediate and small-$\rho$ regions. In these smaller-$\rho$ regions, the NGLs become comparable in size to those observed with $k_t$ clustering as $\rho$ decreases. On the other hand, the CLs contribution is confined to the small-$\rho$ region and is much smaller compared to the NGLs.

In the next subsection, we will perform the resummation of NGLs to all orders.

\subsection{NGLs at all orders}

Let us begin by briefly reviewing the resummation of NGLs and CLs to all orders using the $k_t$ and anti-$k_t$ clustering algorithms within the large-$\Nc$ approximation. This resummation is carried out numerically via the Monte Carlo dipole-evolution program described in refs. \cite{Dasgupta:2001sh,Dasgupta:2002bw}.

\begin{center}
\begin{figure}[ht]
\centering
\includegraphics[width=0.49\textwidth]{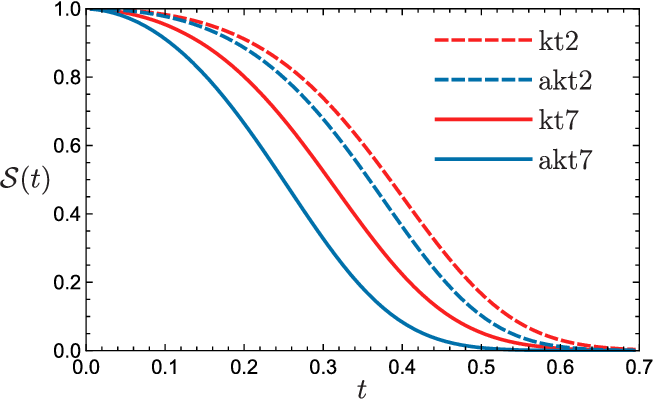}
\includegraphics[width=0.49\textwidth]{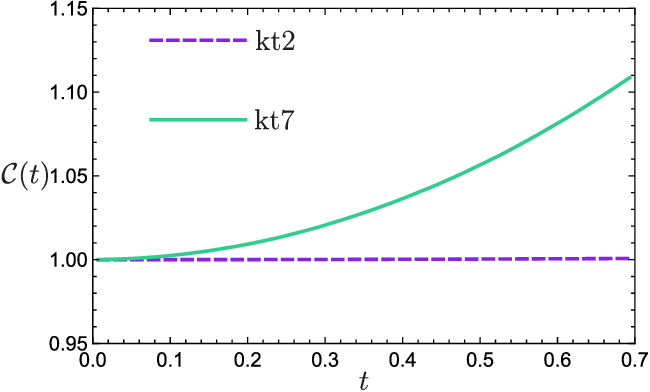}
\caption{\label{fig:03}Resummed NGLs and CLs factors $\mathcal{S}(t)$ and  $\mathcal{C}(t)$ for ungroomed $k_t$ and anti-$k_t$ clustered jets. Results for dipole $(ab)$ are displayed.}
\end{figure}
\end{center}

Figure \ref{fig:03} presents the resummed NGLs factor $\mathcal{S}(t)$ for dipole $(ab)$ using both clustering algorithms with jet radii $R = 0.7$ and $R = 0.2$, alongside the resummed CLs factor $\mathcal{C}(t)$ for $k_t$ clustering, all plotted as a function of the evolution parameter $t$. The evolution parameter $t$ is defined as
\begin{equation}
t(L)=-\frac{1}{4\pi\beta_0}\ln\left(1-2\alpha_s\beta_0L\right),
\end{equation}
where at leading order, $t\to\frac{1}{2}\,\bar{\alpha}_sL$.

One notable observation is that NGLs are reduced when using $k_t$ clustering (with the form factor closer to unity), and they become even smaller for a jet radius of $R = 0.2$. As discussed in the previous subsection, CLs are negligible for $R = 0.2$, with the form factor $\mathcal{C} \sim 1$ for this radius. For instance, at $t = 0.7$ (corresponding to $L \sim 6.78$), the two-loop contributions $f^{\mathrm{trim},(2),\mathrm{CL}}_{\mathcal{B},q}(\rho) \sim 0.0066$ and $f^{\mathrm{trim},(2),\mathrm{CL}}_{\mathcal{B},g}(\rho) \sim 0.0015$ are numerically insignificant.

\begin{center}
\begin{figure}[ht]
\centering
\includegraphics[width=0.495\textwidth]{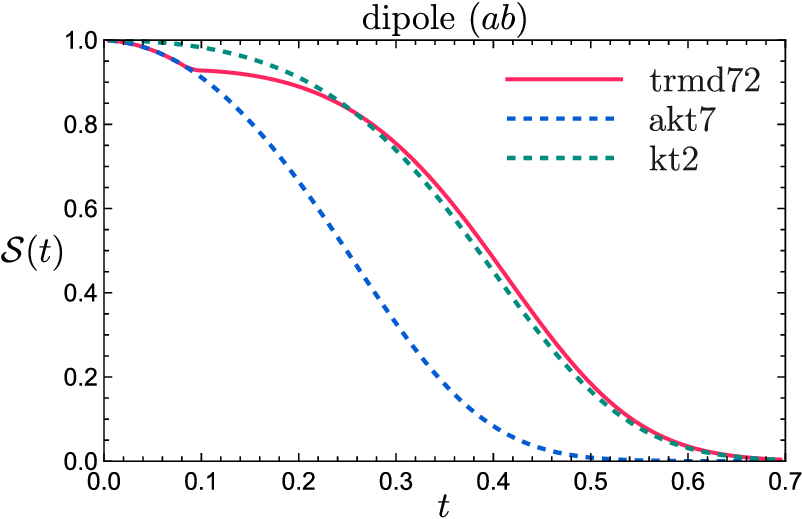}
\includegraphics[width=0.495\textwidth]{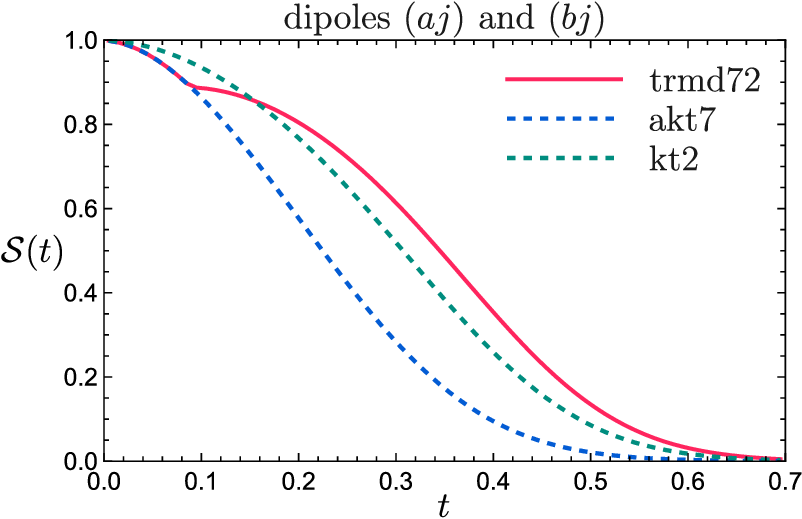}
\caption{\label{fig:02}Resummed NGLs factor $\mathcal{S}(t)$ for trimmed jets. Results for dipole $(ab)$ are displayed on the left, and those for dipoles $(aj)$ and $(bj)$ are shown on the right.}
\end{figure}
\end{center}

For the trimmed distribution, we perform the resummation of NGLs to all orders incorporating the trimming effect in the Monte Carlo program of ref. \cite{Dasgupta:2001sh,Dasgupta:2002bw}. Figure \ref{fig:02} displays the resummed factor $\mathcal{S}(t)$ for different configurations: $R=0.7$ anti-$k_t$ clustering (akt7), $R=0.2$ $k_t$ clustering (kt2), and $R=0.7$ with $R_{\mathrm{sub}}=0.2$ for the trimmed distribution (trmd72). 

In this plot, the trmd72 distribution overlaps with the akt7 distribution at small $t$ (corresponding to large $\rho$). However, for $t>t_{\mathrm{cut}}=t(L_{\mathrm{cut}})$, with $L_{\mathrm{cut}}=\ln(1/z_{\mathrm{cut}})$, the trmd72 distribution departs from the akt7 distribution and eventually approaches the kt2 distribution at large $t$ values.

To incorporate the NGLs into the anti-$k_t$ ungroomed distribution, we parameterize the form factor for each dipole $(i\ell)$ as follows
\begin{align}
\mathcal{S}_{(i\ell)}(t)&=\exp\left(-\frac{\mathcal{C}_{i\ell}}{\CA}\sum_{n=2}^\infty\mathcal{I}^{(n)}_{i\ell}\,\frac{(2\,\CA\,t)^n}{n!}\right),
\end{align}
where the leading-order term is calculated at $\mathcal{O}(\alpha_s^2)$, i.e., $\mathcal{I}^{(2)}_{i\ell}=\mathcal{G}_{2,(i\ell)}(R^2,R^2)$. Numerically we have $\mathcal{I}^{(2)}_{ab}=0.449$ and $\mathcal{I}^{(2)}_{aj}=\mathcal{I}^{(2)}_{bj}=0.823$. The color factor used in the fit is $\mathcal{C}_{i\ell}=2\,\CF$ since the dipole-evolution code assumes a $q\bar{q}$ dipole. In the large-$\Nc$ limit, this factor is set to $2\CF \to \CA$, but for phenomenology, we restore the actual color factors.

The polynomial series in the exponent is truncated once it provides a reasonable fit to the Monte Carlo data. Truncating at $n=5$ offers an accurate description of the Monte Carlo results, with errors at the percent level. The obtained fitting parameters are shown in table \ref{tab:fit}.
\begin{table}
\centering
\begin{tabular}{|c|c|c|c|c|}
  \hline
$n$ & 2 & 3 & 4 & 5\\ \hline 
$\mathcal{I}^{(n)}_{(ab)}$ &  0.449 & 0.153 & 0.048 & 1.242\\ \hline
$\mathcal{I}^{(n)}_{(aj)} $ and $\mathcal{I}^{(n)}_{(bj)}$ & 0.823 & $-0.337$ & 0.380 & 0.429  \\  \hline
$\widetilde{\mathcal{I}}^{(n)}_{(ab)}$ & NA & $-0.624$ & 3.575 & $-4.989$ \\ \hline
$\widetilde{\mathcal{I}}^{(n)}_{(aj)} $ and $\widetilde{\mathcal{I}}^{(n)}_{(bj)}$ & NA & $-1.214$ & 7.795 &  $-12.282$ \\ \hline
\end{tabular}
\caption{The fitting parameters for the Monte Carlo data for the resummed NGLs.\label{tab:fit}}
\end{table}

The obtained coefficients $\mathcal{I}^{(n)}$ are then used in the full non-global form factor, which for the ungroomed distribution is expressed for each channel as
\begin{equation}
\mathcal{S}_{\delta}^{\mathrm{ungr}}(t)=\exp\left(-\sum_{(i\ell)}\frac{\mathcal{C}_{i\ell}}{\CA}\sum_{n=2}\mathcal{I}^{(n)}_{i\ell}\, \frac{(2\,\CA\,t)^n}{n!}\right).
\end{equation}

The trimmed non-global factor can similarly be fitted using the following expression
\begin{align}
\mathcal{S}^{\mathrm{trim}}_{(i\ell)}(t,t_{\mathrm{cut}},t_{\mathrm{sub}})&=\exp\left(f_{\mathcal{B},(i\ell)}^{\mathrm{trim},(2),\mathrm{NG}}(t,t_{\mathrm{cut}},t_\mathrm{sub})-\frac{\mathcal{C}_{i\ell}}{\CA}\sum_{n=3}\mathcal{I}^{(n)}_{i\ell}\, \frac{(2\,\CA\min[t,t_{\mathrm{cut}}])^n}{n!} \right.\notag\\ &\left.-\Theta(t-t_{\mathrm{cut}})\,\frac{\mathcal{C}_{i\ell}}{\CA}\,\sum_{n=3}\widetilde{\mathcal{I}}^{(n)}_{i\ell}\,\frac{(2\,\CA)^n}{n!}(t^n-t_{\mathrm{cut}}^n)\right).
\end{align}
Here, $f_{\mathcal{B},(i\ell)}^{\mathrm{trim},(2),\mathrm{NG}}(t,t_{\mathrm{cut}},t_{\mathrm{sub}})$ is obtained by replacing $\bar{\alpha}_s \ln (R^2/\rho)$ with $2\,t$, $\bar{\alpha}_s \ln (1/z_{\mathrm{cut}})$ with $2\,t_{\mathrm{cut}}$, and $\bar{\alpha}_s \ln (R_{\mathrm{sub}}^2/\rho)$ with $2\,t_{\mathrm{sub}}$ in the expression \eqref{eq:NGres}. This form ensures that the expansion of this resummed factor reproduces the $\mathcal{O}(\alpha_s^2)$ results for NGLs obtained in the previous subsection. The obtained coefficients $\widetilde{\mathcal{I}}^{(n)}_{i\ell}$ are listed in table \ref{tab:fit}.

Since the CLs have a numerically insignificant impact, we neglect their all-orders resummation.

\section{Convolution and numerical results}

In this section, we present the numerical results for the convolution of the resummed form factor, including resummed NGLs at large $\Nc$, with the Born cross-section. We use the Monte Carlo program \texttt{MadGraph5\_aMC@NLO} \cite{Maltoni:2002qb,Alwall:2014hca} to generate Born-level events for the process $pp\to Z+\mathrm{jet}$ at parton level. The generated events are subject to selection cuts $\Omega_\mathcal{B}$, specifically requiring the jet to have $p_t>150$ GeV and $|y|<2.5$. We employ \texttt{NNPDF30\_nlo} parton distribution functions with factorization and renormalization scales set to 150 GeV.

The resulting \texttt{LHE} event sample is accessed with \texttt{MadAnalysis 5} analysis package \cite{Conte:2012fm} to extract the flavor and kinematics of the hard partons in each event. For each bin of the $\rho$ variable, events are weighted with the resummed form factor, and the weights are summed over all events. The integrated distribution, as given in eq. \eqref{eq:master}, is then obtained by dividing the summed weights by the effective luminosity, $\mathcal{L}=N_{\mathrm{tot}}/\sigma_0$, where $N_{\mathrm{tot}}$ is the total number of generated events, and $\sigma_0$ is the Born cross-section obtained from \texttt{MadGraph5\_aMC@NLO} event generator. The differential distribution is calculated through numerical differentiation of the integrated distribution.

\begin{center}
\begin{figure}[ht]
\centering
\includegraphics[width=0.49\textwidth]{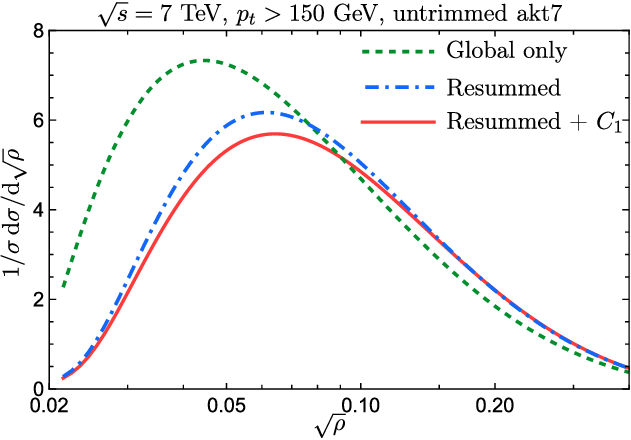}
\includegraphics[width=0.49\textwidth]{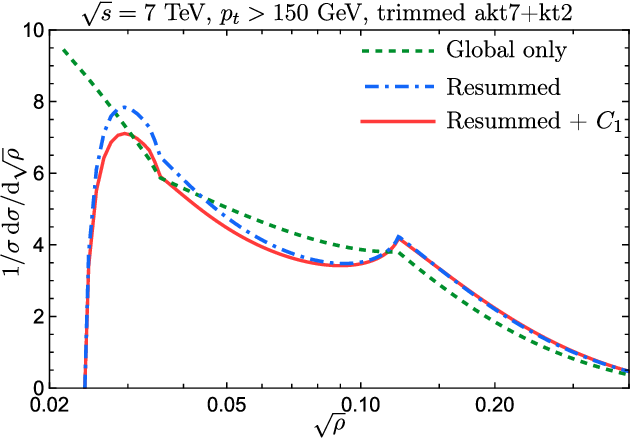}\\
\includegraphics[width=0.49\textwidth]{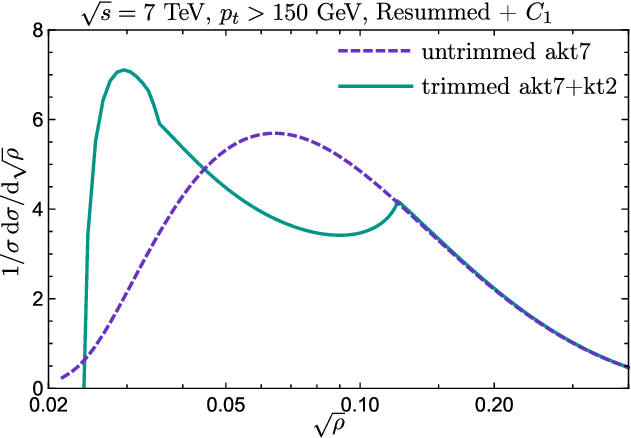}
\caption{\label{fig:07}Resummed differential distributions of the jet mass observable, comparing trimmed and untrimmed cases. The plots display results for global logarithms, global and non-global logarithms, and the full resummation at NLL' accuracy.}
\end{figure}
\end{center}

To illustrate the impact of NGLs and the constant $\alpha_s C_1$, we plot three configurations of the resummed distribution in figure \ref{fig:07}: (a) resumming global logarithms only, (b) resumming both global and non-global logarithms, and (c) the full resummed distribution, valid at NLL' accuracy, including the constant $C_1$. The plots show differential distributions in $\sqrt{\rho}$ normalized to the Born cross-section for the global and resummed distributions, and normalized to the NLO cross-section for the distribution accounting for the constant $C_1$. Results are shown for both ungroomed and trimmed jets.

Our results highlight the significant impact of the non-global contributions on the ungroomed distribution, particularly in the region where large logarithms are dominant. While the non-global effects are less pronounced for the trimmed distribution, they remain non-negligible.

It is important to note that due to the Landau pole singularity, the distribution becomes unreliable for $\rho$ values to the left of the peak, roughly corresponding to $\sqrt{\rho}$ less than $0.03$. This region is also heavily influenced by non-perturbative effects, which are not accounted for in these plots. Additionally, matching to fixed-order NLO results is necessary to correct the behavior of the distribution at the tail, making it comparable to experimental data. We leave these tasks to our future work.

\begin{center}
\begin{figure}[ht]
\centering
\includegraphics[width=0.49\textwidth]{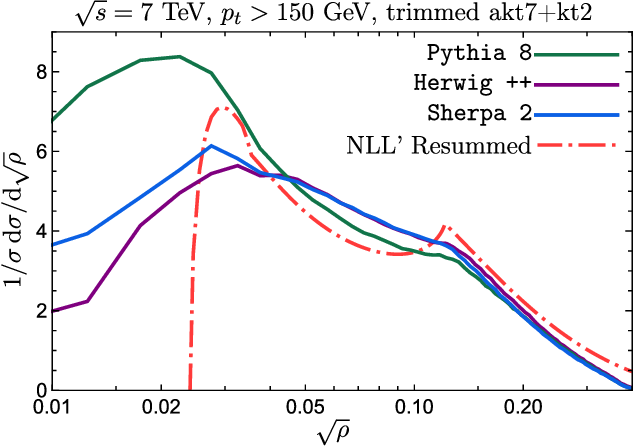}
\includegraphics[width=0.49\textwidth]{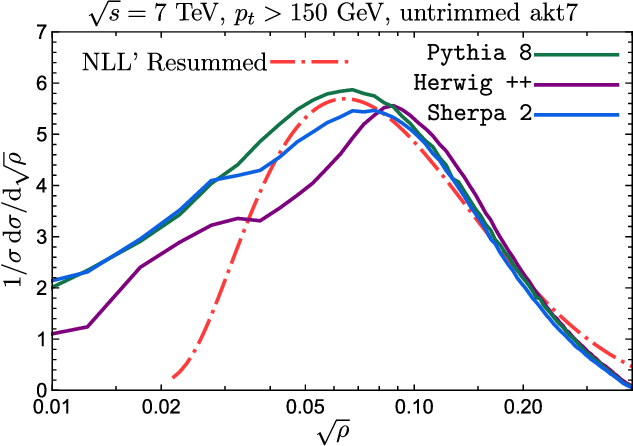}
\caption{\label{fig:08}Comparison of trimmed (left) and untrimmed (right) resummed jet mass distributions with Monte Carlo simulations from parton showers: \texttt{Pythia 8.3}, \texttt{Herwig++}, and \texttt{Sherpa 2.2}.}
\end{figure}
\end{center}

The peak of the untrimmed distribution is more distinct and occurs at a higher mass value than that of the trimmed distribution. This behavior is consistent with the removal of soft and wide-angle radiation from the jet in the trimming process, resulting in a lower overall mass. This trend aligns with the observations from parton shower simulations, as shown in figure \ref{fig:08}.

Figure \ref{fig:08} compares our resummed results, valid at NLL accuracy in the exponent and NLL' accuracy in the expansion, to parton-level Monte Carlo simulations from \texttt{Pythia 8.3} \cite{Bierlich:2022pfr,Alwall:2008qv}, \texttt{Herwig++} \cite{Bahr:2008pv,Bellm:2015jjp}, and \texttt{Sherpa 2.2} \cite{Sherpa:2019gpd}. The \texttt{Pythia 8} and \texttt{Herwig++} results are obtained by showering NLO events from \texttt{MadGraph5\_aMC@NLO}. Jet finding for both ungroomed and groomed jets is performed using \texttt{FastJet} \cite{Cacciari:2011ma} and its plugin \texttt{FastJet-Contrib}.

We observe good agreement between our results and the \texttt{Pythia 8} parton shower, particularly away from the region affected by the Landau pole singularity. For the ungroomed distribution, \texttt{Sherpa} and \texttt{Herwig++} show a distorted peak, particularly to the left of it. As observed in our previous work \cite{Ziani:2021dxr}, this distortion and the discrepancies between different Monte Carlo simulations diminish once non-perturbative effects are included. A similar observation is made for the trimmed distribution, where significant differences between \texttt{Pythia 8} and \texttt{Herwig++}/\texttt{Sherpa} are observed for $\sqrt{\rho}$ values less than 0.05.

\section{Conclusions}

In this paper we have presented a detailed calculation of the resummed trimmed jet mass distribution at NLL accuracy in the exponent and NLL' accuracy in the perturbative expansion of the cumulative distribution. We performed fixed-order calculations at $\mathcal{O}(\alpha_s)$ for global logarithms and $\mathcal{O}(\alpha_s^2)$ for non-global and clustering logarithms, and presented a resummed result valid to all orders. Our $\mathcal{O}(\alpha_s)$ results were found to be consistent with Monte Carlo predictions from the \texttt{MCFM} program.

The distribution exhibits distinct behavior across three key regions of the jet mass variable
\begin{enumerate}
\item  In the large-$\rho$ region, the distribution is unaffected by trimming.
\item 
In the intermediate-$\rho$ region, the double logarithms in $\rho$ vanish and are replaced by constant logarithms of $1/z_{\mathrm{cut}}$. Non-global logarithms in this region become leading, significantly influencing the distribution.
\item 
In the small-$\rho$ region, the double logarithmic structure is restored, and the non-global logarithms converge to those of $k_t$ clustering, characterized by the sub-jet radius $R_{\mathrm{sub}}$. Clustering logarithms appear only in this region, though their numerical impact on the distribution is insignificant.
\end{enumerate}

Following this, we performed a convolution of the resummed form factor, incorporating non-global effects, with the Born cross-section. Our obtained results were compared to parton shower Monte Carlo simulations from \texttt{Pythia 8.3}, \texttt{Herwig++}, and \texttt{Sherpa 2.2} event generators, with non-perturbative effects switched off, yielding reasonable agreement.

While this work focused on obtaining the NLL' resummed distribution, further steps are necessary to make meaningful comparisons with experimental data, such as that provided by the CMS collaboration \cite{CMS:2013kfv}. Specifically, the inclusion of NLO corrections via matching to fixed-order calculations is required. Moreover, non-perturbative corrections, including contributions from the underlying event and hadronization, are expected to have a significant impact on the distribution \cite{Ziani:2021dxr, Dasgupta:2012hg}, and must be included before any comparison with experimental results. These corrections, along with assessments of statistical and systematic uncertainties, will be addressed in our future work.

\acknowledgments

This work is supported by PRFU research project B00L02UN050120230003. We wish to thank the Algerian Ministry of Higher Education and Scientific Research and DGRSDT for financial support. Some of the numerical calculations in this paper have been performed in the High-Performance-Computing cluster at the University of Batna 2 (UB2-HPC). S. Gaid and R. Soualah would like to thank the ICTP for the hospitality, while part of this work was carried out.

\appendix

\section{\label{sec:Rad}Radiator}

The radiator is divided into two contributions, depending on the value of the $\rho$ variable. The total radiator can be expressed as
\begin{equation}\label{eq:rad}
\mathcal{R}_\delta(\rho)=\Theta(\rho-R^2z_{\mathrm{cut}})\,\mathcal{R}^{\mathrm{ungr}}_\delta(\rho,R)+\Theta(R^2z_{\mathrm{cut}}-\rho)\,\mathcal{R}^{\mathrm{trim}}_\delta(\rho,R,R_{\mathrm{sub}},z_{\mathrm{cut}})\,.
\end{equation}

The ungroomed radiator is given by
\begin{equation}
\mathcal{R}_\delta^{\mathrm{ungr}}(\rho,R)=C_j\,\left[L\,g_1(\lambda)+g_2(\lambda)+g_{2,\mathrm{coll}}(\lambda)\right]+g_{2,\mathrm{wide}}(\lambda)\sum_{(i\ell)}\mathcal{C}_{i\ell}\,\mathcal{J}_{i\ell}(R^2)\,,
\end{equation}
where the jet functions $\mathcal{J}_{i\ell}(R^2)$ for the various dipoles are defined in eqs. \eqref{eq:jfab} and \eqref{eq:jfaj}. The LL and NLL resummation functions are given by
\begin{subequations}
\begin{align}
g_1(\lambda)&=\frac{1}{2\pi\beta_0\lambda}\left[(1-2\lambda)\ln(1-2\lambda)-2(1-\lambda)\ln(1-\lambda)\right],\\
g_2(\lambda)&=\frac{\beta_1}{2\pi\beta_0^3}\left[\frac{1}{2}\ln^2(1-2\lambda)+\ln(1-2\lambda)-\ln^2(1-\lambda)-2\ln(1-\lambda)\right]\notag\\
&+\frac{\mathrm{K}}{4\pi^2\beta_0^2}\left[2\ln(1-\lambda)-\ln(1-2\lambda)\right],\\
g_{2,\mathrm{coll}}(\lambda)&=-B_j\,\frac{1}{\pi\beta_0}\,\ln(1-\lambda)\,,\\
g_{2,\mathrm{wide}}(\lambda)&=-\frac{1}{2\pi\beta_0}\ln(1-2\lambda)\,,
\end{align}
\end{subequations}
where $\lambda=\alpha_s(R^2p^2_t)\,\beta_0\,L$, and $L=\ln(R^2/\rho)$. The factor $\mathrm{K}$ accounts for the transition from the Catani-Marchesini-Webber (CMW) scheme \cite{Catani:1990rr} to the $\overline{\text{MS}}$ renormalization scheme
\begin{equation}
\mathrm{K}=\CA\left(\frac{67}{18}-\frac{\pi^2}{6}\right)-\frac{5}{9}\,\mathrm{n_f}\,.\\
\end{equation}
This result includes the two-loop QCD running coupling effect, such that
\begin{equation}
\alpha_s(k_{t}^2)=\alpha_s(p_t^2)\left[\frac{1}{1+2\,\alpha_s\,\beta_0\ln(k_t/p_t)}-\frac{\beta_1}{\beta_0}\,\alpha_s\,\frac{\ln\left[1+2\,\alpha_s\,\beta_0\ln(k_t/p_t)\right]}{[1+2\,\alpha_s\,\beta_0\ln(k_t/p_t)]^2}\right].
\end{equation}
The one- and two-loop coefficients of the QCD beta function are given by
\begin{subequations}
\begin{align}
\beta_0&=\frac{11 \CA-2\mathrm{n_f}}{12 \pi}\,,\\
\beta_1&=\frac{17\mathrm{C_A^2}-5\CA\mathrm{n_f}-3\CF\mathrm{n_f}}{24\pi^2}\,.
\end{align}
\end{subequations}

The trimmed radiator, which modifies the ungroomed radiator in the intermediate and small-$\rho$ regions, is
\begin{align}
\mathcal{R}_{\delta}^{\mathrm{trim}}(\rho,R,R_{\mathrm{sub}},z_{\mathrm{cut}})&=C_j\left[L\,g_1^{\mathrm{trim}}(\lambda,\lambda_{\mathrm{sub}},\lambda_{\mathrm{cut}})+g^{\mathrm{trim}}_2(\lambda,\lambda_{\mathrm{sub}},\lambda_{\mathrm{cut}}) +g_{2,\mathrm{coll}}(\lambda)\right] \notag\\
&+g_{2,\mathrm{wide}}(\lambda)\sum_{(i\ell)}\mathcal{C}_{i\ell}\,\mathcal{J}_{i\ell}(\max[\rho/z_{\mathrm{cut}},R_{\mathrm{sub}}^2])+\notag\\
&+g_{2,\mathrm{wide}}(\lambda_{\mathrm{cut}})\sum_{(i\ell)}\mathcal{C}_{i\ell}\left[\mathcal{J}_{i\ell}(R^2)-\mathcal{J}_{i\ell}(\max[\rho/z_{\mathrm{cut}},R_{\mathrm{sub}}^2])\right],
\end{align}
with $\lambda_{\mathrm{cut}}=\alpha_s(R^2p^2_t)\beta_0L_{\mathrm{cut}}$, $\lambda_{\mathrm{sub}}=\alpha_s(R^2p^2_t)\beta_0L_{\mathrm{sub}}$, $L_{\mathrm{cut}}=\ln(1/z_{\mathrm{cut}})$, and $L_{\mathrm{sub}}=\ln(R_{\mathrm{sub}}^2/\rho)$. The trimming resummation functions are
\begin{subequations}
\begin{align}
g_1^{\mathrm{trim}}&=\frac{1}{2\pi\beta_0\lambda}
\left[\left(1-\lambda-\max[\lambda_{\mathrm{cut}},\lambda_{\mathrm{sub}}]\right)\ln\left(1-\lambda-\max[\lambda_{\mathrm{cut}},\lambda_{\mathrm{sub}}]\right)\right.\notag\\
&+\left(1-\lambda-2\lambda_{\mathrm{cut}}+\max(\lambda_{\mathrm{cut}},\lambda_{\mathrm{sub}})\right)\ln\left(1-\lambda-2\lambda_{\mathrm{cut}}+\max(\lambda_{\mathrm{cut}},\lambda_{\mathrm{sub}})\right)\notag\\
&\left.-2\left(1-\lambda\right)\ln(1-\lambda)-\left(1-2\lambda_{\mathrm{cut}}\right)\ln(1-2\lambda_{\mathrm{cut}})\right],\\
g_2^{\mathrm{trim}}&=\frac{\mathrm{K}}{4\pi^2\beta_0^2}\left[-\ln(1-\lambda-\max[\lambda_{\mathrm{cut}},\lambda_{\mathrm{sub}}])-\ln(1-\lambda-2\lambda_{\mathrm{cut}}+\max[\lambda_{\mathrm{sub}},\lambda_{\mathrm{cut}}])\right.\notag\\
&\left.+2\ln(1-\lambda )+\ln(1-2\lambda_{\mathrm{cut}})\right]\notag\\
&+\frac{\beta_1}{2\pi\beta_0^3}
\left[\frac{1}{2}\ln ^2(1-\lambda -\max[\lambda_{\mathrm{cut}},\lambda_{\mathrm{sub}}])+\frac{1}{2}\ln ^2(1-\lambda -2\lambda_{\mathrm{cut}}+\max[\lambda_{\mathrm{cut}},\lambda_{\mathrm{sub}}])+\notag\right.\\
&\left.+\ln (1-\lambda -\max[\lambda_{\mathrm{cut}},\lambda_{\mathrm{sub}}])+\ln (1-\lambda -2\lambda_{\mathrm{cut}}+\max[\lambda_{\mathrm{cut}},\lambda_{\mathrm{sub}}])-\ln ^2(1-\lambda )\right.\notag\\
&\left.-2 \ln (1-\lambda )-\frac{1}{2} \ln ^2(1-2 \lambda_{\mathrm{cut}})-\ln (1-2 \lambda_{\mathrm{cut}})\right].
\end{align}
\end{subequations}
Note that the radiator is continuous across the transition points.

The derivatives of the ungroomed and trimmed radiators with respect to $L$ and $L_{\mathrm{cut}}$ are given by
\begin{subequations}
\begin{align}
\frac{\partial \mathcal{R}_{\delta}^{\mathrm{ungr}}}{\partial L} &= \frac{C_j}{\pi\beta_0} \left[\ln (1-\lambda)-\ln(1-2\lambda)\right],\\
\frac{\partial \mathcal{R}_{\delta}^{\mathrm{trim}}}{\partial L} &= \frac{C_j}{\pi\beta_0} \left[\ln (1-\lambda)-\ln(1-\lambda-\max[\lambda_{\mathrm{sub}},\lambda_{\mathrm{cut}}])\right],\\
\frac{\partial \mathcal{R}_{\delta}^{\mathrm{trim}}}{\partial L_{\mathrm{cut}}} &= \frac{C_j}{\pi\beta_0} \left[\ln (1-2\lambda_{\mathrm{cut}})-\ln(1-\lambda-2\lambda_{\mathrm{cut}}+\max[\lambda_{\mathrm{cut}},\lambda_{\mathrm{sub}}])\right].
\end{align}
\end{subequations}
Then, the expression for the derivative entering the resummed formula \eqref{eq:master2} is given by
\begin{align}\label{eq:rad2}
\mathcal{R}'_{\delta}(\rho)&\equiv \Theta(\rho-R^2z_{\mathrm{cut}})\,\frac{\partial \mathcal{R}^{\mathrm{ungr}}}{\partial L}+\Theta(R^2z_{\mathrm{cut}}-\rho)\,\left[\frac{\partial \mathcal{R}_{\delta}^{\mathrm{trim}}}{\partial L}+\frac{\partial \mathcal{R}_{\delta}^{\mathrm{trim}}}{\partial L_{\mathrm{cut}}}\right].
\end{align}

\section{\label{sec:CLs}Clustering functions}

The coefficient functions entering the expressions for the CLs factors in eq. \eqref{eq:cff} are given by
\begin{align}
\gamma(x)&  =\frac{1}{32\,\pi}\left[\left(2\,x^2+1\right)\sqrt{4\,x^2-1}+8\,x^2\left(x^2-1\right)\sec^{-1}(2\,x)\right],\\
\mu_2(x) &  =\frac{1}{48\,\pi}\left[6\sqrt{4\,x^2-1}+12\left(2\,x^2-1-2\ln x\right)\sec ^{-1}(2x)+12\,\Im\left(\text{Li}_2[-e^{i\,2\,\sec^{-1}(2 x)}]\right)\right],\\
\mu_4(x) &  =\frac{1}{4}\,\gamma(x)\,,\\
\mu_6(x) &  =\frac{\left(24\,x^4+2\,x^2+1\right) \sqrt{4\,x^2-1}+24\,x^4\left(4\,x^2-3\right)\sec^{-1}(2\,x)}{82\,944\,\pi }\,,\\
\sigma_4(x)&=\frac{1}{4}\,\gamma(x)\,,\\
\sigma_6(x)&=\frac{1}{13\,824\,\pi }\left[\left(6\,x^4+x^2+2\right)\sqrt{4\,x^2-1}+12\,x^2\left(2\,x^4-1\right)\sec^{-1}(2\,x)\right],\\
\tau_6(x)  &=\frac{1}{110\,592\pi}\left[\left(36\,x^4-16\,x^2+7\right)\sqrt{4\,x^2-1}+24\,x^2\left(6\,x^4-x^2-1\right)\sec^{-1}(2\,x)\right],\\
\nu_0(x)&=\tau_0(x)\,,\\
\nu_6(x)&=\frac{1}{331\,776\,\pi}\left[\left(12\,x^4+64\,x^2+5\right)\sqrt{4\,x^2-1}+24\,x^2\left(2\,x^4-3\,x^2-3\right)\sec^{-1}(2\,x)\right].
\end{align}
Here, $x> 1/2$. The remaining functions do not have simple analytical expressions and are shown in table \ref{tab:coef2} for selected values of $x$.
\begin{table}[ht]
\centering
\begin{tabular}{|c|c|c|c|c|c|c|}
  \hline
$x$   & $\sigma_2(x)$ & $\tau_0(x)$ & $\tau_2(x)$ & $\tau_4(x)$ & $\nu_2$ & $\nu_4$  \\ \hline 
$0.5$ &  0.0000 & 0.0000 & 0.0000 & 0.0000  & 0.0000  & 0.000 \\ \hline
$0.6$ &  0.00055& 0.0017 & 0.0011 & 0.00012 &0.00011 & $2.2\times 10^{-6} $ \\ \hline
$0.7$ &  0.0031 & 0.0077 & 0.0057 & 0.00074 &0.00055 & 0.000020\\ \hline
$0.8$ &  0.0087 & 0.0176 & 0.0148 & 0.0022  &0.0014  & 0.000073\\ \hline
$0.9$ &  0.0181 & 0.0305 & 0.0286 & 0.0049  &0.0026  & 0.00019\\ \hline
$1.0$ &  0.0319 & 0.0457 & 0.0475 & 0.0091  &0.0042  & 0.00039\\ \hline
\end{tabular}
\caption{Numerically computed coefficients of the series \eqref{eq:cff} for the $k_t$ CLs factors, evaluated for selected values of $x=R_{\mathrm{sub}}/R$.\label{tab:coef2}}
\end{table}

\bibliographystyle{JHEP}
\bibliography{Refs}

\end{document}